June 2009

# Calibration and Monitoring of the Pierre Auger Observatory

Presentations for the
31st International Cosmic Ray Conference, Łódź , Poland, July 2009

# PIERRE AUGER COLLABORATION

**J. Abraham[8], P. Abreu[71], M. Aglietta[54], C. Aguirre[12], E.J. Ahn[87], D. Allard[31], I. Allekotte[1], J. Allen[90], J. Alvarez-Muñiz[78], M. Ambrosio[48], L. Anchordoqui[104], S. Andringa[71], A. Anzalone[53], C. Aramo[48], E. Arganda[75], S. Argirò[51], K. Arisaka[95], F. Arneodo[55], F. Arqueros[75], T. Asch[38], H. Asorey[1], P. Assis[71], J. Aublin[33], M. Ave[96], G. Avila[10], T. Bäcker[42], D. Badagnani[6], K.B. Barber[11], A.F. Barbosa[14], S.L.C. Barroso[20], B. Baughman[92], P. Bauleo[85], J.J. Beatty[92], T. Beau[31], B.R. Becker[101], K.H. Becker[36], A. Bellétoile[34], J.A. Bellido[11, 93], S. BenZvi[103], C. Berat[34], P. Bernardini[47], X. Bertou[1], P.L. Biermann[39], P. Billoir[33], O. Blanch-Bigas[33], F. Blanco[75], C. Bleve[47], H. Blümer[41, 37], M. Boháčová[96, 27], D. Boncioli[49], C. Bonifazi[33], R. Bonino[54], N. Borodai[69], J. Brack[85], P. Brogueira[71], W.C. Brown[86], R. Bruijn[81], P. Buchholz[42], A. Bueno[77], R.E. Burton[83], N.G. Busca[31], K.S. Caballero-Mora[41], L. Caramete[39], R. Caruso[50], W. Carvalho[17], A. Castellina[54], O. Catalano[53], L. Cazon[96], R. Cester[51], J. Chauvin[34], A. Chiavassa[54], J.A. Chinellato[18], A. Chou[87, 90], J. Chudoba[27], J. Chye[89d], R.W. Clay[11], E. Colombo[2], R. Conceição[71], B. Connolly[102], F. Contreras[9], J. Coppens[65, 67], A. Cordier[32], U. Cotti[63], S. Coutu[93], C.E. Covault[83], A. Creusot[73], A. Criss[93], J. Cronin[96], A. Curutiu[39], S. Dagoret-Campagne[32], R. Dallier[35], K. Daumiller[37], B.R. Dawson[11], R.M. de Almeida[18], M. De Domenico[50], C. De Donato[46], S.J. de Jong[65], G. De La Vega[8], W.J.M. de Mello Junior[18], J.R.T. de Mello Neto[23], I. De Mitri[47], V. de Souza[16], K.D. de Vries[66], G. Decerprit[31], L. del Peral[76], O. Deligny[30], A. Della Selva[48], C. Delle Fratte[49], H. Dembinski[40], C. Di Giulio[49], J.C. Diaz[89], P.N. Diep[105], C. Dobrigkeit [18], J.C. D'Olivo[64], P.N. Dong[105], A. Dorofeev[88], J.C. dos Anjos[14], M.T. Dova[6], D. D'Urso[48], I. Dutan[39], M.A. DuVernois[98], R. Engel[37], M. Erdmann[40], C.O. Escobar[18], A. Etchegoyen[2], P. Facal San Luis[96, 78], H. Falcke[65, 68], G. Farrar[90], A.C. Fauth[18], N. Fazzini[87], F. Ferrer[83], A. Ferrero[2], B. Fick[89], A. Filevich[2], A. Filipčič[72, 73], I. Fleck[42], S. Fliescher[40], C.E. Fracchiolla[85], E.D. Fraenkel[66], W. Fulgione[54], R.F. Gamarra[2], S. Gambetta[44], B. García[8], D. García Gámez[77], D. Garcia-Pinto[75], X. Garrido[37, 32], G. Gelmini[95], H. Gemmeke[38], P.L. Ghia[30, 54], U. Giaccari[47], M. Giller[70], H. Glass[87], L.M. Goggin[104], M.S. Gold[101], G. Golup[1], F. Gomez Albarracin[6], M. Gómez Berisso[1], P. Gonçalves[71], M. Gonçalves do Amaral[24], D. Gonzalez[41], J.G. Gonzalez[77, 88], D. Góra[41, 69], A. Gorgi[54], P. Gouffon[17], S.R. Gozzini[81], E. Grashorn[92], S. Grebe[65], M. Grigat[40], A.F. Grillo[55], Y. Guardincerri[4], F. Guarino[48], G.P. Guedes[19], J. Gutiérrez[76], J.D. Hague[101], V. Halenka[28], P. Hansen[6], D. Harari[1], S. Harmsma[66, 67], J.L. Harton[85], A. Haungs[37], M.D. Healy[95], T. Hebbeker[40], G. Hebrero[76], D. Heck[37], C. Hojvat[87], V.C. Holmes[11], P. Homola[69], J.R. Hörandel[65], A. Horneffer[65], M. Hrabovský[28, 27], T. Huege[37], M. Hussain[73], M. Iarlori[45], A. Insolia[50], F. Ionita[96], A. Italiano[50], S. Jiraskova[65], M. Kaducak[87], K.H. Kampert[36], T. Karova[27], P. Kasper[87], B. Kégl[32], B. Keilhauer[37], E. Kemp[18], R.M. Kieckhafer[89], H.O. Klages[37], M. Kleifges[38], J. Kleinfeller[37], R. Knapik[85], J. Knapp[81], D.-H. Koang[34], A. Krieger[2], O. Krömer[38], D. Kruppke-Hansen[36], F. Kuehn[87], D. Kuempel[36], N. Kunka[38], A. Kusenko[95], G. La Rosa[53], C. Lachaud[31], B.L. Lago[23], P. Lautridou[35], M.S.A.B. Leão[22], D. Lebrun[34], P. Lebrun[87], J. Lee[95], M.A. Leigui de Oliveira[22], A. Lemiere[30], A. Letessier-Selvon[33], M. Leuthold[40], I. Lhenry-Yvon[30], R. López[59], A. Lopez Agüera[78], K. Louedec[32], J. Lozano Bahilo[77], A. Lucero[54], H. Lyberis[30], M.C. Maccarone[53], C. Macolino[45], S. Maldera[54], D. Mandat[27], P. Mantsch[87], A.G. Mariazzi[6], I.C. Maris[41], H.R. Marquez Falcon[63], D. Martello[47], O. Martínez Bravo[59], H.J. Mathes[37], J. Matthews[88, 94], J.A.J. Matthews[101], G. Matthiae[49], D. Maurizio[51], P.O. Mazur[87], M. McEwen[76], R.R. McNeil[88], G. Medina-Tanco[64], M. Melissas[41], D. Melo[51], E. Menichetti[51], A. Menshikov[38], R. Meyhandan[14], M.I. Micheletti[2], G. Miele[48], W. Miller[101], L. Miramonti[46], S. Mollerach[1], M. Monasor[75], D. Monnier Ragaigne[32], F. Montanet[34], B. Morales[64], C. Morello[54], J.C. Moreno[6], C. Morris[92], M. Mostafá[85], C.A. Moura[48], S. Mueller[37], M.A. Muller[18], R. Mussa[51], G. Navarra[54], J.L. Navarro[77], S. Navas[77], P. Necesal[27], L. Nellen[64], C. Newman-Holmes[87], D. Newton[81], P.T. Nhung[105], N. Nierstenhoefer[36], D. Nitz[89], D. Nosek[26], L. Nožka[27], M. Nyklicek[27], J. Oehlschläger[37], A. Olinto[96], P. Oliva[36], V.M. Olmos-Gilbaja[78], M. Ortiz[75], N. Pacheco[76], D. Pakk Selmi-Dei[18], M. Palatka[27], J. Pallotta[3], G. Parente[78], E. Parizot[31], S. Parlati[55], S. Pastor[74], M. Patel[81], T. Paul[91], V. Pavlidou[96c], K. Payet[34], M. Pech[27], J. Pękala[69], I.M. Pepe[21], L. Perrone[52], R. Pesce[44], E. Petermann[100], S. Petrera[45], P. Petrinca[49], A. Petrolini[44], Y. Petrov[85], J. Petrovic[67], C. Pfendner[103], R. Piegaia[4], T. Pierog[37], M. Pimenta[71], T. Pinto[74], V. Pirronello[50], O. Pisanti[48], M. Platino[2], J. Pochon[1], V.H. Ponce[1], M. Pontz[42], P. Privitera[96], M. Prouza[27], E.J. Quel[3], J. Rautenberg[36], O. Ravel[35],**

**D. Ravignani[2], A. Redondo[76], B. Revenu[35], F.A.S. Rezende[14], J. Ridky[27], S. Riggi[50], M. Risse[36], C. Rivière[34], V. Rizi[45], C. Robledo[59], G. Rodriguez[49], J. Rodriguez Martino[50], J. Rodriguez Rojo[9], I. Rodriguez-Cabo[78], M.D. Rodríguez-Frías[76], G. Ros[75], [76], J. Rosado[75], T. Rossler[28], M. Roth[37], B. Rouillé-d'Orfeuil[31], E. Roulet[1], A.C. Rovero[7], F. Salamida[45], H. Salazar[59b], G. Salina[49], F. Sánchez[64], M. Santander[9], C.E. Santo[71], E.M. Santos[23], F. Sarazin[84], S. Sarkar[79], R. Sato[9], N. Scharf[40], V. Scherini[36], H. Schieler[37], P. Schiffer[40], A. Schmidt[38], F. Schmidt[96], T. Schmidt[41], O. Scholten[66], H. Schoorlemmer[65], J. Schovancova[27], P. Schovánek[27], F. Schroeder[37], S. Schulte[40], F. Schüssler[37], D. Schuster[84], S.J. Sciutto[6], M. Scuderi[50], A. Segreto[53], D. Semikoz[31], M. Settimo[47], R.C. Shellard[14], [15], I. Sidelnik[2], B.B. Siffert[23], A. Śmiałkowski[70], R. Šmída[27], B.E. Smith[81], G.R. Snow[100], P. Sommers[93], J. Sorokin[11], H. Spinka[82], [87], R. Squartini[9], E. Strazzeri[32], A. Stutz[34], F. Suarez[2], T. Suomijärvi[30], A.D. Supanitsky[64], M.S. Sutherland[92], J. Swain[91], Z. Szadkowski[70], A. Tamashiro[7], A. Tamburro[41], T. Tarutina[6], O. Taşcău[36], R. Tcaciuc[42], D. Tcherniakhovski[38], D. Tegolo[58], N.T. Thao[105], D. Thomas[85], R. Ticona[13], J. Tiffenberg[4], C. Timmermans[67], [65], W. Tkaczyk[70], C.J. Todero Peixoto[22], B. Tomé[71], A. Tonachini[51], I. Torres[59], P. Travnicek[27], D.B. Tridapalli[17], G. Tristram[31], E. Trovato[50], M. Tueros[6], R. Ulrich[37], M. Unger[37], M. Urban[32], J.F. Valdés Galicia[64], I. Valiño[37], L. Valore[48], A.M. van den Berg[66], J.R. Vázquez[75], R.A. Vázquez[78], D. Veberič[73], [72], A. Velarde[13], T. Venters[96], V. Verzi[49], M. Videla[8], L. Villaseñor[63], S. Vorobiov[73], L. Voyvodic[87‡], H. Wahlberg[6], P. Wahrlich[11], O. Wainberg[2], D. Warner[85], A.A. Watson[81], S. Westerhoff[103], B.J. Whelan[11], G. Wieczorek[70], L. Wiencke[84], B. Wilczyńska[69], H. Wilczyński[69], C. Wileman[81], M.G. Winnick[11], H. Wu[32], B. Wundheiler[2], T. Yamamoto[96a], P. Younk[85], G. Yuan[88], A. Yushkov[48], E. Zas[78], D. Zavrtanik[73], [72], M. Zavrtanik[72], [73], I. Zaw[90], A. Zepeda[60b], M. Ziolkowski[42]**

*[1] Centro Atómico Bariloche and Instituto Balseiro (CNEA-UNCuyo-CONICET), San Carlos de Bariloche, Argentina*
*[2] Centro Atómico Constituyentes (Comisión Nacional de Energía Atómica/CONICET/UTN- FRBA), Buenos Aires, Argentina*
*[3] Centro de Investigaciones en Láseres y Aplicaciones, CITEFA and CONICET, Argentina*
*[4] Departamento de Física, FCEyN, Universidad de Buenos Aires y CONICET, Argentina*
*[6] IFLP, Universidad Nacional de La Plata and CONICET, La Plata, Argentina*
*[7] Instituto de Astronomía y Física del Espacio (CONICET), Buenos Aires, Argentina*
*[8] National Technological University, Faculty Mendoza (CONICET/CNEA), Mendoza, Argentina*
*[9] Pierre Auger Southern Observatory, Malargüe, Argentina*
*[10] Pierre Auger Southern Observatory and Comisión Nacional de Energía Atómica, Malargüe, Argentina*
*[11] University of Adelaide, Adelaide, S.A., Australia*
*[12] Universidad Catolica de Bolivia, La Paz, Bolivia*
*[13] Universidad Mayor de San Andrés, Bolivia*
*[14] Centro Brasileiro de Pesquisas Fisicas, Rio de Janeiro, RJ, Brazil*
*[15] Pontifícia Universidade Católica, Rio de Janeiro, RJ, Brazil*
*[16] Universidade de São Paulo, Instituto de Física, São Carlos, SP, Brazil*
*[17] Universidade de São Paulo, Instituto de Física, São Paulo, SP, Brazil*
*[18] Universidade Estadual de Campinas, IFGW, Campinas, SP, Brazil*
*[19] Universidade Estadual de Feira de Santana, Brazil*
*[20] Universidade Estadual do Sudoeste da Bahia, Vitoria da Conquista, BA, Brazil*
*[21] Universidade Federal da Bahia, Salvador, BA, Brazil*
*[22] Universidade Federal do ABC, Santo André, SP, Brazil*
*[23] Universidade Federal do Rio de Janeiro, Instituto de Física, Rio de Janeiro, RJ, Brazil*
*[24] Universidade Federal Fluminense, Instituto de Fisica, Niterói, RJ, Brazil*
*[26] Charles University, Faculty of Mathematics and Physics, Institute of Particle and Nuclear Physics, Prague, Czech Republic*
*[27] Institute of Physics of the Academy of Sciences of the Czech Republic, Prague, Czech Republic*
*[28] Palacký University, Olomouc, Czech Republic*
*[30] Institut de Physique Nucléaire d'Orsay (IPNO), Université Paris 11, CNRS-IN2P3, Orsay, France*
*[31] Laboratoire AstroParticule et Cosmologie (APC), Université Paris 7, CNRS-IN2P3, Paris, France*
*[32] Laboratoire de l'Accélérateur Linéaire (LAL), Université Paris 11, CNRS-IN2P3, Orsay, France*
*[33] Laboratoire de Physique Nucléaire et de Hautes Energies (LPNHE), Universités Paris 6 et Paris 7, Paris Cedex 05, France*

[34] *Laboratoire de Physique Subatomique et de Cosmologie (LPSC), Université Joseph Fourier, INPG, CNRS-IN2P3, Grenoble, France*
[35] *SUBATECH, Nantes, France*
[36] *Bergische Universität Wuppertal, Wuppertal, Germany*
[37] *Forschungszentrum Karlsruhe, Institut für Kernphysik, Karlsruhe, Germany*
[38] *Forschungszentrum Karlsruhe, Institut für Prozessdatenverarbeitung und Elektronik, Karlsruhe, Germany*
[39] *Max-Planck-Institut für Radioastronomie, Bonn, Germany*
[40] *RWTH Aachen University, III. Physikalisches Institut A, Aachen, Germany*
[41] *Universität Karlsruhe (TH), Institut für Experimentelle Kernphysik (IEKP), Karlsruhe, Germany*
[42] *Universität Siegen, Siegen, Germany*
[44] *Dipartimento di Fisica dell'Università and INFN, Genova, Italy*
[45] *Università dell'Aquila and INFN, L'Aquila, Italy*
[46] *Università di Milano and Sezione INFN, Milan, Italy*
[47] *Dipartimento di Fisica dell'Università del Salento and Sezione INFN, Lecce, Italy*
[48] *Università di Napoli "Federico II" and Sezione INFN, Napoli, Italy*
[49] *Università di Roma II "Tor Vergata" and Sezione INFN, Roma, Italy*
[50] *Università di Catania and Sezione INFN, Catania, Italy*
[51] *Università di Torino and Sezione INFN, Torino, Italy*
[52] *Dipartimento di Ingegneria dell'Innovazione dell'Università del Salento and Sezione INFN, Lecce, Italy*
[53] *Istituto di Astrofisica Spaziale e Fisica Cosmica di Palermo (INAF), Palermo, Italy*
[54] *Istituto di Fisica dello Spazio Interplanetario (INAF), Università di Torino and Sezione INFN, Torino, Italy*
[55] *INFN, Laboratori Nazionali del Gran Sasso, Assergi (L'Aquila), Italy*
[58] *Università di Palermo and Sezione INFN, Catania, Italy*
[59] *Benemérita Universidad Autónoma de Puebla, Puebla, Mexico*
[60] *Centro de Investigación y de Estudios Avanzados del IPN (CINVESTAV), México, D.F., Mexico*
[61] *Instituto Nacional de Astrofisica, Optica y Electronica, Tonantzintla, Puebla, Mexico*
[63] *Universidad Michoacana de San Nicolas de Hidalgo, Morelia, Michoacan, Mexico*
[64] *Universidad Nacional Autonoma de Mexico, Mexico, D.F., Mexico*
[65] *IMAPP, Radboud University, Nijmegen, Netherlands*
[66] *Kernfysisch Versneller Instituut, University of Groningen, Groningen, Netherlands*
[67] *NIKHEF, Amsterdam, Netherlands*
[68] *ASTRON, Dwingeloo, Netherlands*
[69] *Institute of Nuclear Physics PAN, Krakow, Poland*
[70] *University of Łódź, Łódź, Poland*
[71] *LIP and Instituto Superior Técnico, Lisboa, Portugal*
[72] *J. Stefan Institute, Ljubljana, Slovenia*
[73] *Laboratory for Astroparticle Physics, University of Nova Gorica, Slovenia*
[74] *Instituto de Física Corpuscular, CSIC-Universitat de València, Valencia, Spain*
[75] *Universidad Complutense de Madrid, Madrid, Spain*
[76] *Universidad de Alcalá, Alcalá de Henares (Madrid), Spain*
[77] *Universidad de Granada & C.A.F.P.E., Granada, Spain*
[78] *Universidad de Santiago de Compostela, Spain*
[79] *Rudolf Peierls Centre for Theoretical Physics, University of Oxford, Oxford, United Kingdom*
[81] *School of Physics and Astronomy, University of Leeds, United Kingdom*
[82] *Argonne National Laboratory, Argonne, IL, USA*
[83] *Case Western Reserve University, Cleveland, OH, USA*
[84] *Colorado School of Mines, Golden, CO, USA*
[85] *Colorado State University, Fort Collins, CO, USA*
[86] *Colorado State University, Pueblo, CO, USA*
[87] *Fermilab, Batavia, IL, USA*
[88] *Louisiana State University, Baton Rouge, LA, USA*
[89] *Michigan Technological University, Houghton, MI, USA*
[90] *New York University, New York, NY, USA*
[91] *Northeastern University, Boston, MA, USA*
[92] *Ohio State University, Columbus, OH, USA*
[93] *Pennsylvania State University, University Park, PA, USA*
[94] *Southern University, Baton Rouge, LA, USA*
[95] *University of California, Los Angeles, CA, USA*

[96] *University of Chicago, Enrico Fermi Institute, Chicago, IL, USA*
[98] *University of Hawaii, Honolulu, HI, USA*
[100] *University of Nebraska, Lincoln, NE, USA*
[101] *University of New Mexico, Albuquerque, NM, USA*
[102] *University of Pennsylvania, Philadelphia, PA, USA*
[103] *University of Wisconsin, Madison, WI, USA*
[104] *University of Wisconsin, Milwaukee, WI, USA*
[105] *Institute for Nuclear Science and Technology (INST), Hanoi, Vietnam*
‡ *Deceased*
[a] *at Konan University, Kobe, Japan*
[b] *On leave of absence at the Instituto Nacional de Astrofisica, Optica y Electronica*
[c] *at Caltech, Pasadena, USA*
[d] *at Hawaii Pacific University*

# The monitoring system of the Pierre Auger Observatory and its additional functionalities

**J. Rautenberg*** **for the Pierre Auger Collaboration**†

*Bergische Universität Wuppertal, 42097 Wuppertal, Germany
†Av. San Martin Norte 304 (5613) Malargüe, Prov. de Mendoza, Argentina

***Abstract*****. To ensure smooth operation of the Pierre Auger Observatory a monitoring tool has been developed. Data from different sources, e.g. the detector components, are collected and stored in a single database. The shift crew and experts can access these data using a web interface that displays generated graphs and specially developed visualisations. This tool offers an opportunity to monitor the long term stability of some key quantities and of the data quality. Quantities derived such as the on-time of the fluorescence telescopes can be estimated in nearly real-time and added to the database for further analysis. In addition to access via the database server the database content is distributed in packages allowing a wide range of analysis off-site. A new functionality has been implemented to manage maintenance and intervention in the field using the web interface. It covers the full work-flow from an alarm being raised to the issue being resolved.**

## I. Introduction

The Pierre Auger Observatory is measuring cosmic rays at the highest energies. The southern site in Mendoza, Argentina, has been completed during the year 2008. The instrument [1] has been designed to measure extensive air showers with energies ranging from $10^{18}$ – $10^{20}$ eV and beyond. It combines two complementary observational techniques, the detection of particles on the ground using an array of 1600 water Cherenkov detectors distributed on an area of 3000 km$^2$ and the observation of fluorescence light generated in the atmosphere above the ground by 24 wide-angle Schmidt telescopes positioned in four buildings on the border around the ground array. Routine operation of the detectors has started in 2002. The observatory became fully operational in 2008 with the completion of the construction.

## II. Online-monitoring for the Pierre Auger Observatory

For the optimal scientific output of the observatory the status of the detector as well as its measured data have to be monitored. The Auger Monitoring tool [2] has been developed to support the shifter in judging and supervising the status of the detector components, the electronics and the data-acquisition.

The detector components are operated differently and therefore the monitoring of their status have different requirements. The stations of the surface detector (SD) operate constantly in an semi-automated mode. Data acquisition must be monitored and failures of stations or of their communication must be detected. The data-taking of the fluorescence detector (FD) can only take place under specific environmental conditions and is organized in shifts. The sensitive cameras can only be operated in dark nights with not too strong wind and without rain. This makes the operation a busy task for the shifters that have to judge the operation-mode on the basis of the information given.

The basis of the monitoring system is a database running at the central campus. The front-end is web based using common technologies like PHP, CSS and JavaScript. An interface has been developed for the generation of visualisations. Alarms for situations that require immediate action are first filled into a specified table of the database that is checked by the web front-end.

The content of the database is mirrored on a server in Europe using the MySQL built-in replication mechanism. This way not only the shifter and maintenance staff on site can use the monitoring, but it is also available for experts all around the world.

## III. Higher level quantities implementation

The data collected in the database can be used to derive higher level quantities such as the up-time of the FD telescopes. This quantity is of major importance since it is a necessary ingredient of flux measurements. The dead-time of each telescopes is recorded in the database. Together with the run information and other corrections retrieved from the database the total uptime for each telescope can be determined individually. The up-time is calculated only for time-intervals of ten minutes, balancing the statistical precision of the calculated up-time due to statistics with the information frequency. A program to execute the calculation is running on the database server and fills continuously the appropriate tables in the database. The web-interface displays the stored quantities. An example of one night of data-taking is given in Fig. 1. The up-time is available in quasi real-time for the shifter as a diagnostic and figure of merit.

## IV. Database distribution for analysis use

The information collected in the database is a valuable source for analysis, i.e. for studies of the long term stability of the detector. With increasing measurement time the accumulated data is too large for the online usage by the shifter, who usually focuses on the most recent data with high priority on the immediate response of the monitoring system. On the other hand, for analysis usually only a small part of the database is used. Therefore the database is split into monthly pieces containing only single components of the database. These pieces can be transported and used off-site for analysis. Only the most recent portion of the database is retained for use by the online monitoring system, keeping the response of the system fast for the shifter.

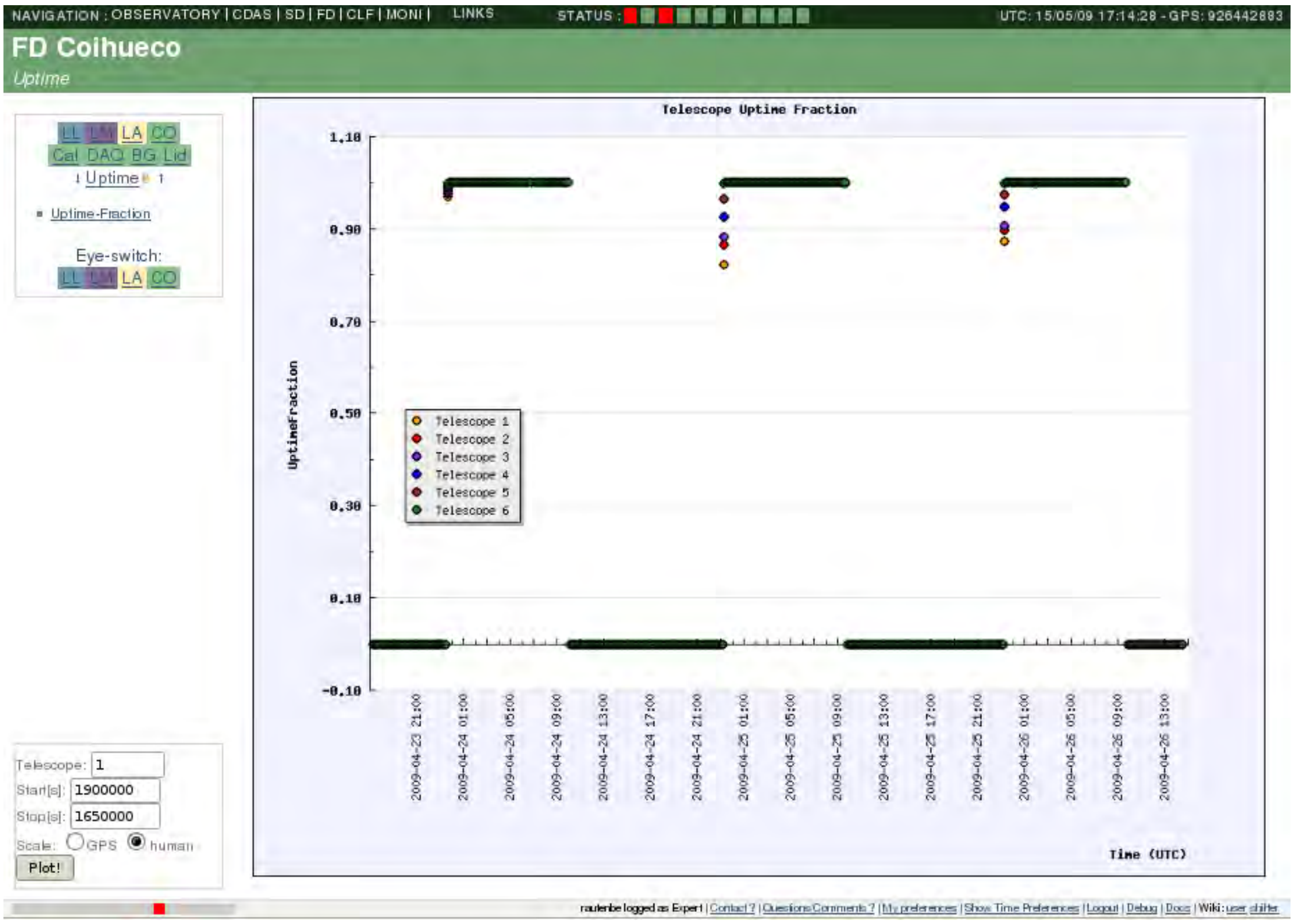

Fig. 1. Example of the web-interface for the display of the uptime. Shown is the uptime-fraction for the six telescopes of one FD building and three nights data-taking.

## V. Maintenance and Intervention Management

For the operation of the surface detector an additional web feature has been developed to manage hardware maintenance and operations in the field triggered by alarms. This is a new part added to the SD section of the Auger Online Monitoring web site and covers the full work-flow from an alarm being raised to being resolved.

In order to check all the components of a SD station various sensors are installed in each station. The calibration process runs online every minute and the sensor measurements and the calibration data are sent to the central data acquisition server every six minutes. These data are transferred to the monitoring database. Analysis software checks the database contents once per day to detect long-term problems such as PMT instabilities, discharging batteries, etc. and fills the database alarm table. The dedicated SD alarm web page allows the shifter to view the alarm table in a user friendly way. Since the table can be huge when all alarms are displayed, selection tools are also provided. They allow to look for a particular set of alarms. From an alarm table link, the shifter can easily have access to the web page of the particular SD station, where he can look at different plots in order to judge the reliability of the alarm. An example for a web page displaying an alarm is given in Fig. 2. From the same page the shifter can consult the history of all alarms that previously occurred on the station and view all the maintenance done or planed for it. Tools are also provided to plot any monitoring or calibration data of the SD station as a function of time.

When a shifter notices an alarm on the Auger Online Monitoring web site, he performs some analysis to check if it is a real alarm requiring an action. He may write a summary into a file and add illustrative plots. Then he contacts the SD Scientific Operation Coordinator (SOC) and SD experts by sending them an email, using the contact link displayed on the footer of every Auger Online Monitoring web page. The link displays a new web page allowing the shifter to enter the the email subject, a comment via a text area, and optional files to upload. After reception of the email, the SD SOC or SD experts can either click on the link automatically added to the email to create a new maintenance request with fields filled out automatically, or connect to the website and add one or more maintenance requests as displayed in Fig. 3. The SD SOC or an SD expert associates then one or several predefined actions to the maintenance request. They can create a mask for one or several types of alarms from the maintenance web page, so that corresponding alarms will not appear by default in the shifter pages.

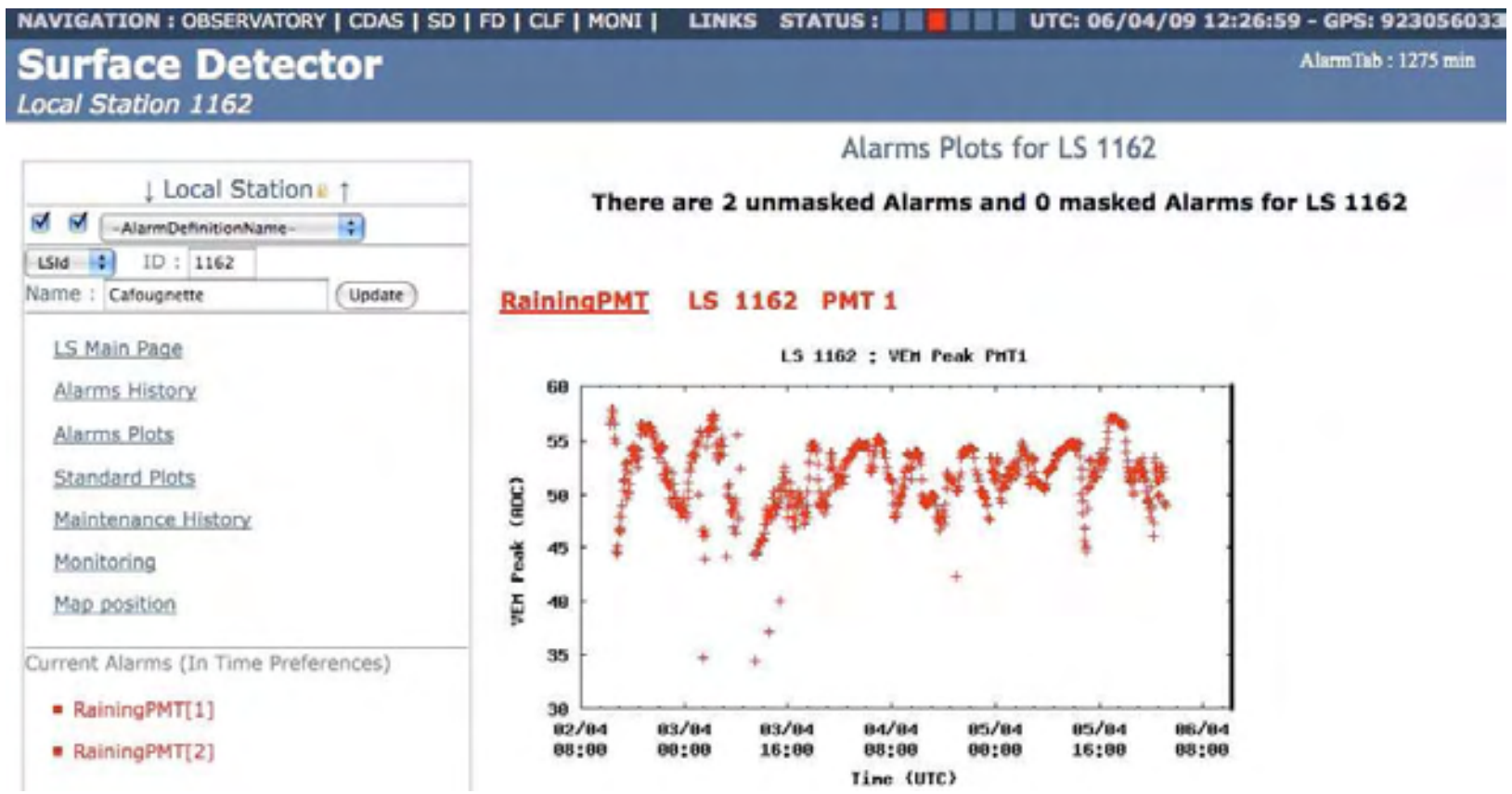

Fig. 2. Example of the web-interface for the display of an SD station alarm plot.

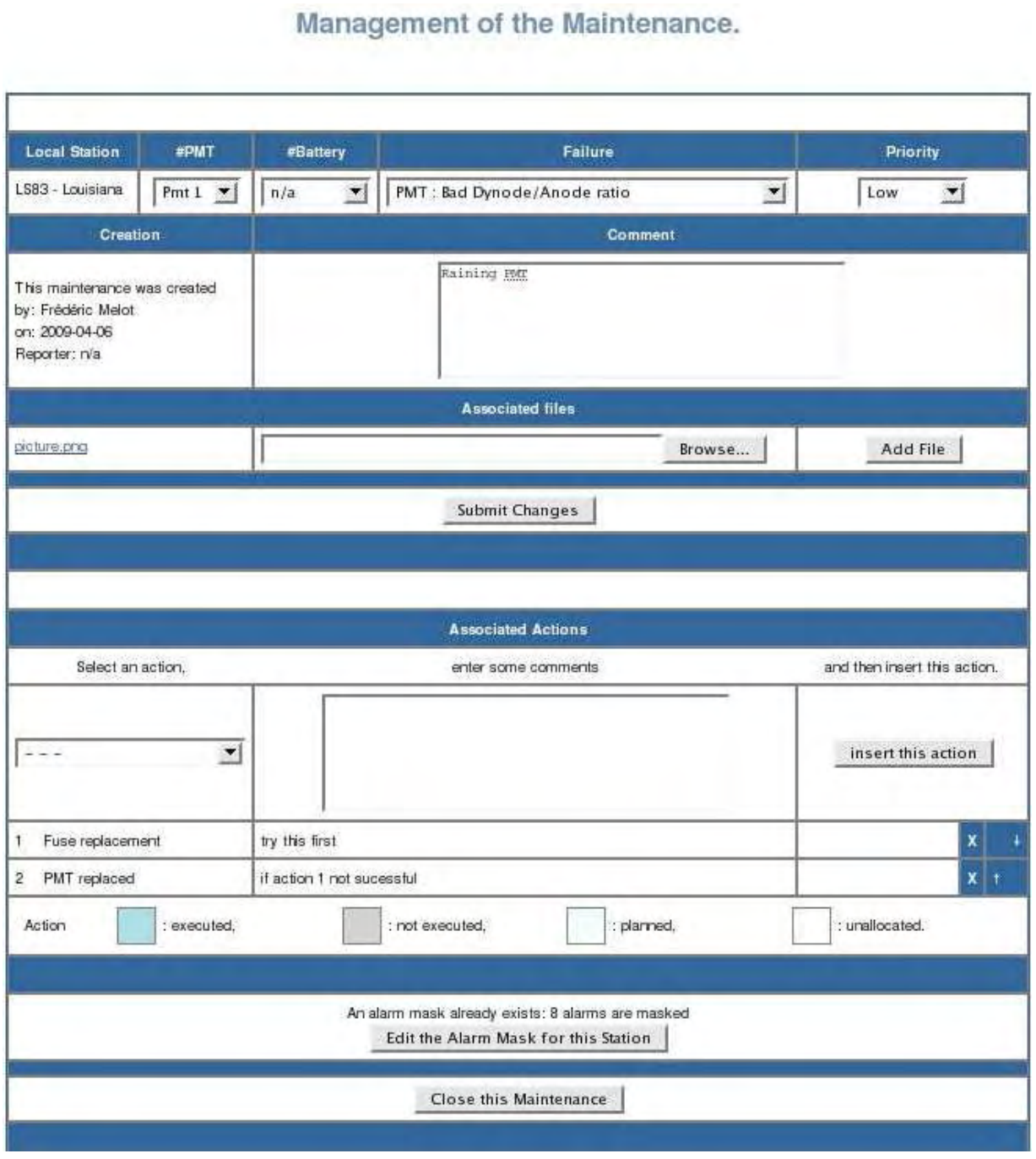

Fig. 3. The web interface available to manage a maintenance.

The task of the SD SOC consists of planning one or several interventions and associating maintenance actions to them. The web interface has been designed to provide the necessary tools to help him. A predefined road map is provided that describes what has to be done during the intervention in order to help the maintenance crew in their work. Returning from the intervention, a maintenance-crew member fills in the actions actually done using the intervention report web page. Doing this "closes" the intervention. Actions not executed can be planned again in another intervention. When no pending action exists on the maintenance, the SD SOC or an SD expert can close the concerned maintenance request, which automatically unmasks the associated masked alarms.

All maintenance, actions and interventions are stored in database tables. Specific web pages with associated tools have been designed in order to view and manage maintenance and interventions.

## VI. SUMMARY

A monitoring tool has been developed for the Pierre Auger Observatory to ensure the quality of the recorded data. Data from different sources are collected and stored in a database accessible for the shift crew and experts via a web interface that displays generated graphs and specially developed visualisations. Higher level quantities such as the on-time of the fluorescence telescopes are derived in nearly real-time and added to the database for further analysis. In addition to access via the database server the database content is distributed in packages allowing a wide range of analysis off-site. A new functionality has been implemented to manage maintenance and intervention in the field using the web interface. It covers the full work-flow from an alarm being raised to the issue being resolved.

# Atmospheric Monitoring and its Use in Air Shower Analysis at the Pierre Auger Observatory

**Segev BenZvi*** **for the Pierre Auger Collaboration**†

*University of Wisconsin – Madison, 222 W. Washington Ave., Madison, WI 53703, USA
†Observatorio Pierre Auger, Av. San Martin Norte 304, 5613 Malargüe, Argentina

***Abstract*. For the analysis of air showers measured using the air fluorescence technique, it is essential to understand the behaviour of the atmosphere. At the Pierre Auger Observatory, the atmospheric properties that affect the production of UV light in air showers and the transmission of the light to the fluorescence telescopes are monitored regularly. These properties include the temperature, pressure, and humidity as a function of altitude; the optical depth and scattering behaviour of aerosols; and the presence of clouds in the field of view of the telescopes. The atmospheric measurements made at the observatory describe a detector volume in excess of 30,000 cubic km. Since 2004, the data have been compiled in a record of nightly conditions, and this record is vital to the analysis of events observed by the fluorescence telescopes. We will review the atmospheric monitoring techniques used at the observatory and discuss the influence of atmospheric measurements on estimates of shower observables using real and simulated data.**

## I. Introduction

The Pierre Auger Observatory comprises two cosmic ray extensive air shower detectors: a Surface Detector Array (SD) of 1600 water Cherenkov detectors, and a Fluorescence Detector (FD) of 24 telescopes at four sites overlooking the array. Observations carried out with the FD yield nearly calorimetric measurements of the energy of each primary cosmic ray. The FD telescopes are also used to observe the slant depth of shower maximum ($X_{\rm max}$), which is sensitive to the mass composition of cosmic rays. Simultaneous shower measurements with the SD and FD (hybrid events) provide high-quality data used in physics analysis and in the calibration of the SD energy scale.

The crucial roles of hybrid calibration and shower measurement performed by the FD depend on detailed knowledge of atmospheric conditions. Light from extensive air showers is produced in the atmosphere, and it is transmitted through the air to the observing telescopes. The production of fluorescence and Cherenkov photons in a shower depends on the temperature, pressure, and humidity of the air. Moreover, as the light travels from the shower axis to the fluorescence telescopes, it is scattered from its path by molecules and aerosols. Therefore, atmospheric conditions have a major impact on shower energies and shower maxima estimated using the fluorescence technique.

To characterise the behaviour of the atmosphere at the Pierre Auger Observatory, extensive atmospheric monitoring is performed during and between FD shifts. Fig. 1 depicts the instruments used in the monitoring program. Atmospheric state variables such as pressure, temperature, and humidity are recorded using meteorological radio soundings launched from a helium balloon station [1], and conditions at ground level are recorded by five weather stations. Aerosol conditions are measured using central lasers, lidars, and cloud cameras [2], [3], [4], as well as optical telescopes and phase function monitors [5], [6]. The atmospheric data have been incorporated into a multi-gigabyte database used for the reconstruction and analysis of hybrid events. We describe the use of these data in estimates of shower light production (Section II) and atmospheric transmission (Section III), and in Section IV we summarise systematic uncertainties in the hybrid reconstruction.

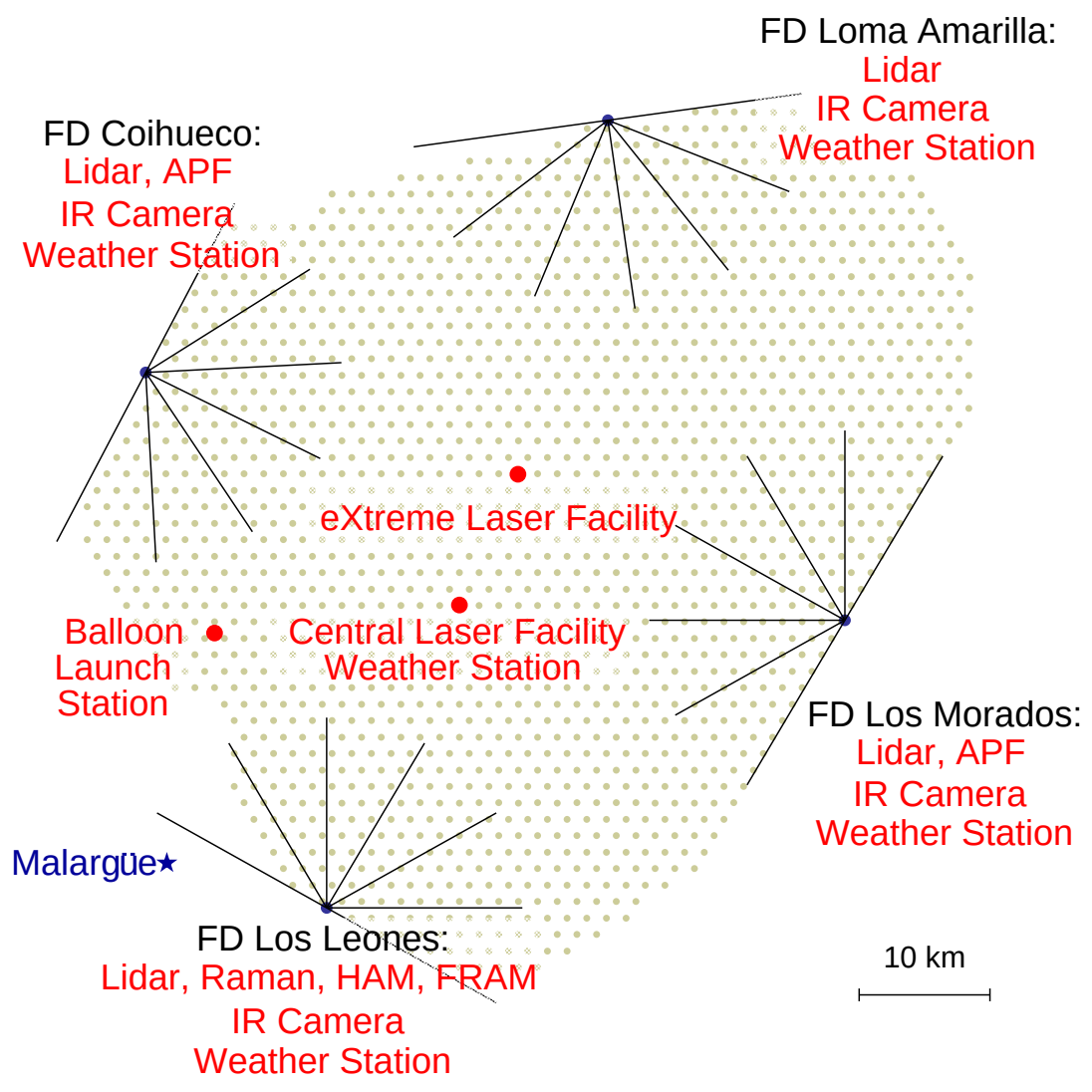

Fig. 1: Atmospheric monitors at the Pierre Auger Observatory include two central lasers, four elastic lidar stations, one Raman lidar, four IR cameras, five weather stations, a balloon launch facility, two aerosol phase function (APF) monitors, and two optical telescopes (HAM, FRAM).

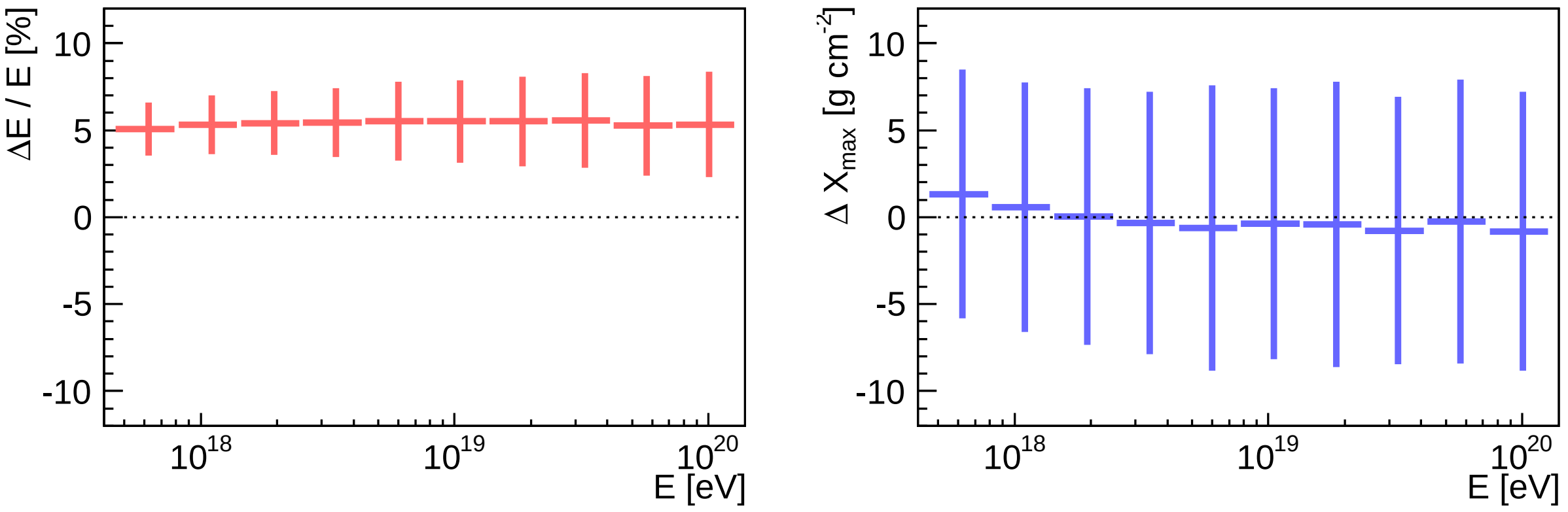

Fig. 2: Combined effects of collisional quenching and atmospheric variability. *Left:* Comparison of reconstructed energies of simulated showers using fluorescence models from AIRFLY [7] and Keilhauer et al. [8]. The uncertainties refer to RMS variations. The AIRFLY model was applied using monthly averages for $p$ and $T$, constant collisional cross sections, and no water vapour quenching; the Keilhauer model was applied using balloon data, $T$-dependent collisional cross sections, and water vapour quenching. *Right:* Comparison of $X_{\rm max}$ using the two models.

## II. Ultraviolet Light Production

Cherenkov and fluorescence production at a given wavelength $\lambda$ depend on the pressure $p$, temperature $T$, and vapour pressure $e$ of the air. The Cherenkov light yield can be determined from the index of refraction of air, but the weather-dependence of the fluorescence yield is considerably more difficult to calculate. This is due to quenching of the radiative transitions of excited $N_2$ by collisions between $N_2$ molecules, collisions between $N_2$ and $O_2$, and collisions between $N_2$ and water vapour. The collisional cross sections depend on temperature and must be determined experimentally [7], [9]. Estimates of these effects in the field are further complicated by significant daily and seasonal variability in the concentration of water vapour.

In Malargüe, the altitude dependence of air pressure, temperature, and relative humidity are measured up to about 23 km above sea level using balloon-borne radio soundings. Balloon launches are performed roughly every five days, and as of this writing there have been 287 successful launches since 2003. Due to the limited measurement statistics, the balloon data were used to create monthly models of atmospheric state variables for use in shower analysis. The models, first introduced in 2005, have recently been updated to include more radiosonde data and humidity profiles [1].

The use of monthly models in the reconstruction provides a significant reduction in systematic uncertainties with respect to the use of a global, static atmospheric model. For example, reconstructing air showers with the 1976 U.S. Standard Atmosphere rather than local monthly models shifts $X_{\rm max}$ by 15 g cm$^{-2}$, on average [10]. When the monthly models are used, a small systematic uncertainty remains due to the daily variability of the atmosphere. We estimate the size of the effect using simulated proton and iron showers between $10^{17.7}$ and $10^{20}$ eV. The showers were reconstructed with monthly average profiles and compared to a reconstruction using 109 cloud-free radio soundings. We find that the monthly models introduce minor systematic shifts into the reconstructed energy ($\Delta E/E = -0.5\%$) and shower maximum ($\Delta X_{\rm max} = 2$ g cm$^{-2}$).

More significant systematic shifts are caused by the collisional cross sections $\sigma_{\rm NN}(T)$ and $\sigma_{\rm NO}(T)$ and the water vapour cross section $\sigma(e)$. We have calculated the effect using simulated showers with UV light generated according to fluorescence models published by AIRFLY [7] and Keilhauer et al. [8]. The addition of $T$-dependent collisional cross sections and water vapour quenching systematically increases the energy by 5.5% and decreases $X_{\rm max}$ by 2 g cm$^{-2}$, partially offsetting the uncertainties due to atmospheric variability. The combined effects of quenching and variability are shown in Fig. 2. The uncertainties RMS($\Delta E/E$)= $1.5\%-3.0\%$ and RMS($\Delta X_{\rm max}$)= $7.2-8.4$ g cm$^{-2}$, which increase in the energy range $10^{17.7}-10^{20}$ eV, are caused by the variability of atmospheric conditions.

## III. Ultraviolet Transmission

When the light from an air shower travels to an FD telescope, it is absorbed and scattered by molecules and aerosols. The attenuation of light is given by the optical transmittance $\mathcal{T}$. In a horizontally uniform atmosphere, the transmittance from an altitude $h$ to the ground through a slanted path of elevation $\varphi$ is

$$\mathcal{T}(h,\lambda,\varphi) = e^{-\tau(h,\lambda)/\sin\varphi}\cdot(1+H.O.), \qquad (1)$$

where the exponential term is the Beer-Lambert law, $\tau(\lambda,h)$ is the total vertical optical depth between the ground and altitude $h$, and $H.O.$ is a higher-order single and multiple scattering correction. The total optical depth is simply the sum of the molecular and aerosol optical depths, which must be either estimated or measured.

### A. Molecular Attenuation

In the lower atmosphere, the attenuation of near-UV light by molecules is predominantly due to scattering. Hence, the vertical molecular optical depth between the ground and altitude $h$ can be calculated from

$$\tau_m(\lambda, h) = \int_{h_{\text{gnd}}}^{h} N(h')\ \sigma_R(\lambda, h')\ dh', \qquad (2)$$

where $N$ is the number density of scatterers and $\sigma_R$ is the Rayleigh scattering cross section [11]. The altitude profiles of pressure, temperature, and vapour pressure can be used to calculate $N(h)$ and $\sigma_R(\lambda, h)$; hence, data from radio soundings or monthly average profiles completely describe molecular scattering. Transmission uncertainties due to the use of monthly models are included in the values reported in Section II.

### B. Aerosol Attenuation

Aerosol attenuation does not have a general analytical solution, and so knowledge of aerosol transmission requires direct field measurements of the aerosol optical depth. To estimate the transmission, we assume the form

$$\tau_a(\lambda, h) = \tau_a(\lambda_0, h) \cdot (\lambda_0/\lambda)^{\gamma}, \qquad (3)$$

where $\tau_a(\lambda_0, h)$ is the vertical aerosol optical depth profile recorded at a single wavelength $\lambda_0$, and the wavelength dependence of $\tau_a(\lambda, h)$ is parameterised by the exponent $\gamma$ [6], [12]. Hourly measurements of the vertical aerosol optical depth profile are carried out using two central lasers [2] ($\lambda_0 = 355$ nm) and four lidar stations [3] ($\lambda_0 = 351$ nm). As shown in Table I, more than 13 000 site-hours of aerosol data have been collected since 2004 using the Central Laser Facility. The data are required inputs to the hybrid physics analysis, and roughly 80% of hybrid events can be reconstructed using aerosol measurements.

We have propagated the measurement uncertainties in the hourly aerosol data into the reconstruction of real hybrid events observed since 2004. Over the energy range $10^{17.7} - 10^{20}$ eV, the average systematic uncertainties in energy increase from $\Delta E/E = ^{+3.6}_{-3.0}\%$ to $^{+7.9}_{-7.0}\%$, and the uncertainties in $X_{\text{max}}$ increase from $\Delta X_{\text{max}}$ $= ^{+3.3}_{-1.3}$ g cm$^{-2}$ to $^{+7.3}_{-4.8}$ g cm$^{-2}$. For the RMS, we make the preliminary estimates RMS($\Delta E/E$)=1.6$(1 \pm 1)\%$ to $2.5(1 \pm 1)\%$ and RMS($\Delta X_{\text{max}}$)=3.0$(1 \pm 1)$ g cm$^{-2}$ to $4.7(1 \pm 1)$ g cm$^{-2}$. The uncertainties are dominated by the aerosol optical depth, with minor contributions from the exponent $\gamma$ and the aerosol phase function. The use of hourly aerosol data offers a significant improvement over a static aerosol model, which if used would increase the systematic uncertainties by a factor of two.

Small horizontal nonuniformities in the vertical aerosol distribution also introduce energy-dependent uncertainties into the reconstruction. The contribution to the average uncertainties is negligible, but over the same energy range, we estimate the effect of the uniformity to be RMS($\Delta E/E$)= $3.6\% - 7.4\%$ and RMS($\Delta X_{\text{max}}$)= $5.7 - 7.6$ g cm$^{-2}$.

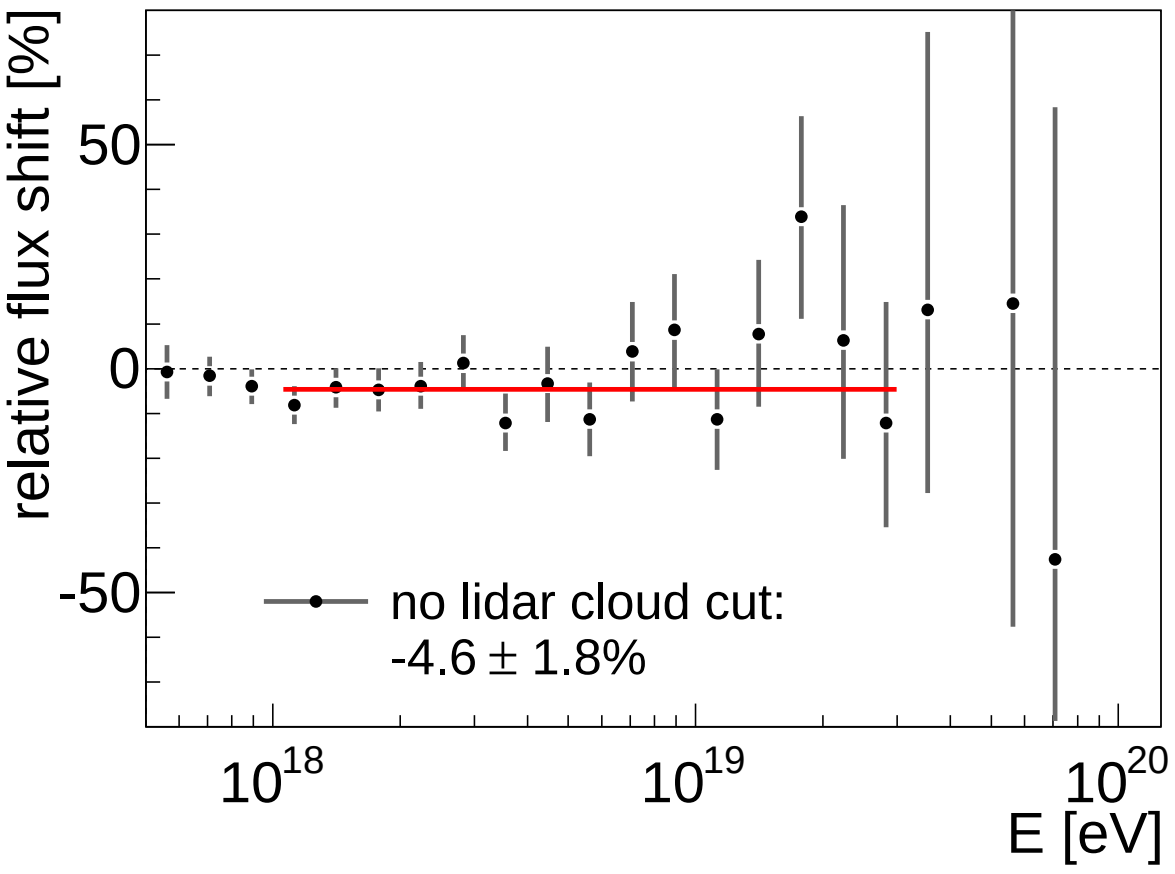

Fig. 3: Relative shift in the flux of events detected in hybrid mode when the lidar cloud coverage cut is not applied. The shift between $10^{18}$ and $10^{19.5}$ eV is indicated by a solid line.

### C. Multiple Scattering Corrections

While molecular and aerosol scattering primarily attenuate shower light as it propagates to an FD telescope, it will also increase the detected signal by scattering photons into the telescope. This causes a systematic overestimate of the shower signal, particularly at low altitudes where the density of scatterers is greatest.

Several Monte Carlo studies have been carried out to parameterise the multiply-scattered component of shower light as a function of optical depth [13], [14]. Using real hybrid events, we have found that a failure to account for multiple scattering will cause overestimates of $2\% - 5\%$ in shower energies and $1 - 3$ g cm$^{-2}$ in $X_{\text{max}}$, where the overestimates increase with energy. Once multiple scattering is included in the reconstruction, the systematic differences between various multiple scattering parameterisations are $\Delta E/E < 1\%$ and $\Delta X_{\text{max}}$ $\approx 1$ g cm$^{-2}$ for all energies.

### D. Attenuation by Clouds

Clouds strongly attenuate UV light, and therefore have a major influence on FD measurements. By blocking the line of sight to high altitudes, cloud layers can bias a fluorescence telescope toward the detection of deep showers and alter the effective aperture of the FD. To monitor clouds and correct for these effects, cloud layers are tracked using the elastic lidar stations and four IR cloud cameras [3], [4]. Data from these instruments are stored in an hourly database of cloud height and coverage above each FD site, up to an altitude of 12 km.

The lidar cloud database has been used to analyse the cloud conditions in Malargüe [3], and the data indicate clear conditions during $50\%$ of measured hours, and $< 25\%$ coverage in $60\%$ of measured hours. The remaining hours are affected by moderate to heavy cloud coverage; $> 80\%$ sky coverage occurs in $20\%$ of the lidar measurements.

TABLE I: Statistics of hourly cloud and aerosol measurements collected at the Pierre Auger Observatory and analysed as of this writing.

| **Aerosol and Cloud Measurements at the Pierre Auger Observatory (2004–2009)** | | | |
|---|---|---|---|
| Location: | Los Leones | Los Morados | Coihueco |
| Aerosols (CLF): | 4943 hours<br>15 Jan 2004 – 5 Mar 2009 | 3760 hours<br>18 Mar 2005 – 5 Mar 2009 | 4695 hours<br>16 Jun 2004 – 5 Mar 2009 |
| Clouds (Lidar): | 3784 hours<br>4 Apr 2006 – 4 Feb 2009 | 3308 hours<br>1 Jul 2006 – 4 Feb 2009 | 4461 hours<br>1 Nov 2005 – 4 Feb 2009 |
| Clouds (IR Cameras): | 4432 hours<br>to May 2008 | 2681 hours<br>to Jan 2008 | 4420 hours<br>to Aug 2008 |

TABLE II: Systematic uncertainties in the hybrid reconstruction due to atmospheric influences on light production and transmission.

| **Systematic Uncertainties** | | | | | |
|---|---|---|---|---|---|
| Source | $\log(E/\text{eV})$ | $\Delta E/E$ (%) | RMS($\Delta E/E$) (%) | $\Delta X_{\text{max}}$ (g cm$^{-2}$) | RMS($\Delta X_{\text{max}}$) (g cm$^{-2}$) |
| *Molecular Light Transmission and Production* | | | | | |
| Horiz. Uniformity | $17.7-20.0$ | 1 | 1 | 1 | 2 |
| Quenching Effects<br>$p$, $T$, $e$ Variability | $17.7-20.0$ | +5.5<br>-0.5 | $1.5-3.0$ | -2.0<br>+2.0 | $7.2-8.4$ |
| *Aerosol Light Transmission* | | | | | |
| $\tau_a(\lambda_0, h)$ | $<18.0$<br>$18.0-19.0$<br>$19.0-20.0$ | $+3.6, -3.0$<br>$+5.1, -4.4$<br>$+7.9, -7.0$ | $1.6\pm1.6$<br>$1.8\pm1.8$<br>$2.5\pm2.5$ | $+3.3, -1.3$<br>$+4.9, -2.8$<br>$+7.3, -4.8$ | $3.0\pm3.0$<br>$3.7\pm3.7$<br>$4.7\pm4.7$ |
| $\gamma$ Exponent | $17.7-20.0$ | 0.5 | 2.0 | 0.5 | 2.0 |
| Phase Function | $17.7-20.0$ | 1.0 | 2.0 | 2.0 | 2.5 |
| Horiz. Uniformity | $<18.0$<br>$18.0-19.0$<br>$19.0-20.0$ | 0.3<br>0.4<br>0.2 | 3.6<br>5.4<br>7.4 | 0.1<br>0.1<br>0.4 | 5.7<br>7.0<br>7.6 |
| *Scattering Corrections* | | | | | |
| Mult. Scattering | $<18.0$<br>$18.0-19.0$<br>$19.0-20.0$ | 0.4<br>0.5<br>1.0 | 0.6<br>0.7<br>0.8 | 1.0<br>1.0<br>1.2 | 0.8<br>0.9<br>1.1 |

The number of hybrid events affected by cloud obscuration is reduced with strong cuts on the shape of reconstructed shower profiles. Showers must also be reconstructed with an hourly aerosol profile from the Central Laser Facility, weighting the data toward periods with relatively unobstructed views to the center of the SD. For the surviving events, a cut of $< 25\%$ lidar cloud coverage has been applied and compared to the dataset with no lidar cuts (Fig. 3). Over the energy range $10^{18}$ to $10^{19.5}$ eV, a $4\%$ reduction is observed in the flux if no cloud cut is applied. Clouds also increase measurements of $\langle X_{\text{max}}\rangle$ by blocking the upper part of the FD fiducial volume; without the lidar cloud cut, we find a systematic increase in $\langle X_{\text{max}}\rangle$ of 3 g cm$^{-2}$ at all energies.

## IV. Summary of Systematic Uncertainties

The Pierre Auger Observatory has accumulated a large database of atmospheric measurements relevant to the production of light in air showers and the transmission of the light to fluorescence telescopes. We have propagated the uncertainties of the atmospheric data into the reconstruction, and estimated the size of effects such as collisional quenching and multiple scattering. The systematic uncertainties are summarised in Table II.

Aside from large "quenching effects" on the fluorescence yield, the uncertainties are dominated by the variability of the molecular atmosphere and the uniformity and uncertainties of the aerosol optical depth. The combined uncertainties are, approximately, $\Delta E/E \approx 4\% - 8\%$, RMS($\Delta E/E$)$\approx 5\pm1\%$ to $9\pm1\%$, $\Delta X_{\text{max}} \approx 4-8$ g cm$^{-2}$, and RMS($\Delta X_{\text{max}}$)$\approx 11\pm1$ g cm$^{-2}$ to $13\pm1$ g cm$^{-2}$. The atmospheric data provide a significant improvement over static weather models, reducing the systematic uncertainties by approximately a factor of two.

# Atmospheric effects on extensive air showers observed with the array of surface detectors of the Pierre Auger Observatory

**Benjamin Rouillé d'Orfeuil*** **for the Pierre Auger Collaboration†**

**Laboratoire AstroParticule et Cosmologie, Université Paris 7, CNRS-IN2P3*
*†Observatorio Pierre Auger, Av. San Martin Norte 304, (5613) Malargüe, Mendoza, Argentina*

***Abstract*****. Atmospheric parameters, such as pressure (P), temperature (T) and density ($\rho \propto$ P/T), affect the development of extensive air showers (EAS) initiated by energetic cosmic rays. We have studied the impact of atmospheric variations on EAS with data from the array of surface detectors of the Pierre Auger Observatory, analysing the dependence of the event rate on P and $\rho$. We show that the observed behaviour is explained by a model including P and $\rho$ and validated with full EAS simulations. Changes in the atmosphere affect also the measured signal, with an impact on the determination of the energy of the primary particle. We show how the energy estimation can be corrected for such effects.**

## I. Introduction

High-energy cosmic rays (CRs) are detected by means of the extensive air shower (EAS) they produce in the atmosphere. The atmosphere affects the EAS development. The properties of the primary CR, such as its energy, have to be inferred from EAS. Therefore the study and understanding of the effects of atmospheric variations on EAS in general, and on a specific detector in particular, is very important for the comprehension of the detector performances and for the correct interpretation of EAS measurements.

We have studied the impact of atmospheric variations on EAS with data collected during 4 years with the array of surface detectors (SD) of the Pierre Auger Observatory, located in Malargüe, Argentina. The Pierre Auger Observatory is designed to study CRs from $\approx 10^{18}$ eV up to the highest energies. The SD consists of 1600 water-Cherenkov detectors to detect the photons and the charged particle of the EAS. It is laid out over 3000 km$^2$ on a triangular grid of 1.5 km spacing and is overlooked by 24 fluorescence telescopes (FD) grouped in units of 6 at four locations on its periphery. For each event, the signals in the stations are fitted to find the signal at a 1000 m core distance, S(1000), which is used to estimate the primary energy.

## II. Impact of atmospheric effects on EAS and their measurement

The water-Cherenkov detectors are sensitive to both the electromagnetic (e.m) component and the muonic component of the EAS, which are influenced to a different extent by atmospheric variations. These in turn influence the signal measured in the detectors and in particular S(1000) [1]. Pressure (P) and air density ($\rho$) are the properties of the atmosphere that affect EAS the most. P changes are associated to changes in the column density of the air above the detector, and hence affect the age of the EAS when they reach the ground. $\rho$ changes modify the Molière radius ($r_M$) and thus influence the lateral attenuation of the EAS. The impact on S(1000) can then be modeled with a Gaisser-Hillas and Nishimura-Kamata-Greisen profile, which describe respectively the longitudinal and the lateral distribution of the e.m component of the EAS. In fact, the relevant value of $r_M$ is the one corresponding to the air density two radiation lengths ($X_0$) above ground in the direction of the incoming EAS [2]. Due to the thermal coupling of the lower atmosphere with the Earth surface, the variation of $\rho$ at $2X_0$ is the same as at the ground on large time scales, while it is smaller on shorter time intervals. It is then useful to separate the dependence of the total signal $S = S_{em} + S_\mu$ on $\rho$ in two terms describing respectively its longer term modulation and its daily one. Introducing the average daily density $\rho_d$ and the instantaneous departure from it, $\rho - \rho_d$, we have:

$$S = S_0 \left[1 + \alpha_P (P - P_0) + \alpha_\rho (\rho_d - \rho_0) + \beta_\rho (\rho - \rho_d)\right] \quad (1)$$

where $S_0$ is the signal that would have been measured at some reference atmospheric conditions with pressure $P_0$ and density $\rho_0$.

The fraction of the signal at 1 km of the core due to the e.m particles is taken as $F_{em} = F_0 - 0.5(\sec\theta - 1)$ with $F_0 = 0.65 + 0.035 \log(E/\mathrm{EeV})$ that provides a reasonable fit to the results of proton EAS simulated for zenith angle $\theta < 60°$ and energies $E = 10^{18}$ to $10^{19}$ eV. The P correlation coefficient is:

$$\alpha_P \simeq -\frac{F_{em}}{g}\left[1 - \frac{\hat{X}_m}{X}\right]\frac{\sec\theta}{\Lambda}$$

where $X = X_v \sec\theta$ is the slant depth with $X_v =$ 880 g cm$^{-2}$ the grammage at the detector site. $\Lambda$ is an effective attenuation length associated to the longitudinal development of the EAS at 1 km from their core and $g$ is the acceleration of gravity. The depth of the EAS maximum at 1 km from the core is $\hat{X}_m \simeq X_m + 150$ g cm$^{-2}$, with $X_m \simeq [700 + 55\log(E/\mathrm{EeV})]$ g cm$^{-2}$ being the average value of the EAS maximum at the core measured by the FD [4]. Due to the flat longitudinal development of the muons, no significant P dependence is expected for the muonic component. The $\rho$ correlation coefficient describing the daily averaged modulation of S is:

$$\alpha_\rho \simeq F_{em}\,\alpha_\rho^{em} + (1 - F_{em})\,\alpha_\rho^{\mu}$$

with:

$$\alpha_\rho^{em} = -\frac{4.5 - 2s}{\rho_0}$$

where $s = 3/(1 + 2\cos\theta X_m/X_v)$ is the shower age. $\alpha_\rho^{\mu}$ is found to be consistent with a zero value in the proton EAS simulations. Concerning the modulation on short time scale, we adopt $\beta_\rho = F_{em}\,\beta_\rho^{em}$ with:

$$\beta_\rho^{em} = \exp(-a\cos\theta)\,\alpha_\rho^{em}$$

where $a$ characterises the amplitude of the daily $\rho$ variation in the lower atmosphere and is completely independent of the EAS development.

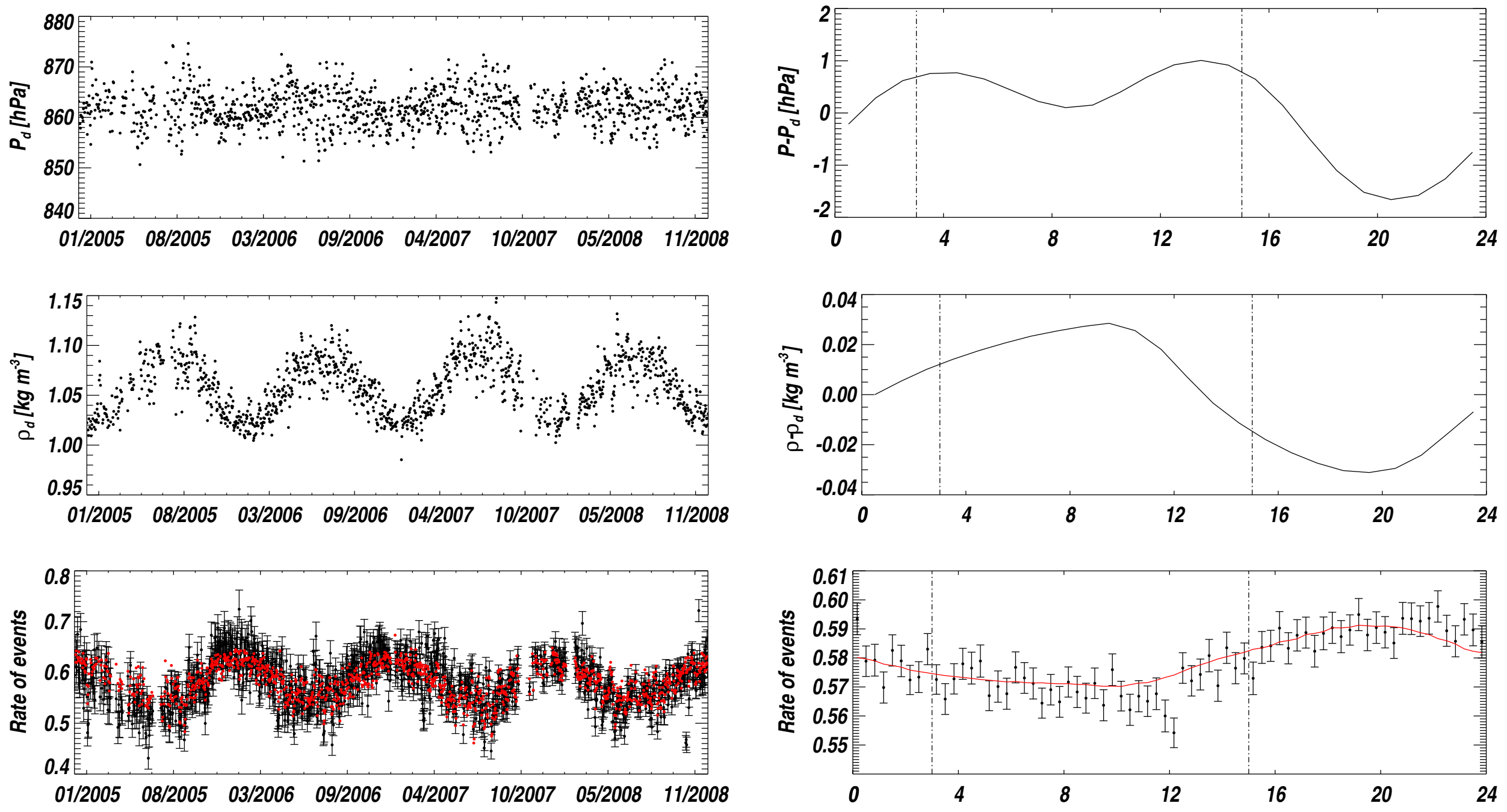

Fig. 1. Left: daily averages of ground P (top), $\rho$ (middle) and rate of events (bottom,black). The prominent effect on the modulation of the rate of events is due to $\rho$ variations. The red points in the bottom plot show the results of the fit. Right: variation of P (top), $\rho$ (middle) and the rate of events during the day (UTC). The vertical dashed lines show the local midnight and noon (UTC-3h) and the red line in the bottom plot show the result of the fit.

As atmospheric variations correspond to signal variations, this implies that the same primary CR will induce different signals depending on P and $\rho$. It follows that the rate of events observed in a given range of S(1000) will be modulated in time. The effect can be quantified starting from the relation between S(1000) and the reconstructed energy: $E_r \propto [S(1000)]^B$, where $B = 1.08{\pm}0.01(\mathrm{stat}){\pm}0.04(\mathrm{sys})$ [3]. Following eq. (1), the primary energy $E_0(\theta, P, \rho)$ that would have been obtained for the same EAS at the reference atmospheric conditions is related to $E_r$ as follows:

$$E_0 = E_r\,[1 - \alpha_\xi \Delta\xi]^B \qquad (2)$$

where $\alpha_\xi \Delta\xi \equiv \alpha_P(P - P_0) + \alpha_\rho(\rho_d - \rho_0) + \beta_\rho(\rho - \rho_d)$. If we focus on a given $\theta$ bin, the rate of events per unit time in a given signal range, $[S_m, S_M]$ is:

$$R(S_m, S_M) = \int_{S_m}^{S_M} \mathrm{d}S\,A(S)\,\frac{\mathrm{d}J}{\mathrm{d}S}$$

where J is the flux of CRs and $A(S)$ is the instantaneous acceptance of the experiment. It will be of the form $A(S) = \kappa\,\epsilon(S)$, where $\kappa$ is a constant global factor proportional to the area of the SD and the solid angle considered, while $\epsilon(S)$ is the trigger probability. Assuming that the CR spectrum is a pure power law, *i.e* $\mathrm{dJ}/\mathrm{dE}_0 \propto \mathrm{E}_0^{-\gamma}$, and using eq. (2) and neglecting the small energy dependence of the coefficients $\alpha_\xi$, we can derive the corresponding dependence of the rate of events:

$$R(S_m, S_M) \propto (1 + a_\xi \Delta\xi) \int_{S_m}^{S_M} \mathrm{d}S\,\epsilon(S)\,S^{-B\gamma + B - 1} \qquad (3)$$

with the coefficients modulating the rate of events being $a_\xi \Delta\xi = B(\gamma - 1)\alpha_\xi \Delta\xi$. This expression implies that for any given values of $S_m$ and $S_M$, the associated rate of events will have the same modulation, regardless of whether the acceptance is saturated ($\epsilon(S) = 1$) or not.

## III. Modulation of the experimental rate of events

To study the expected modulation of the rate of events, we use data taken by the SD from 1 January 2005 to 31 December 2008 with $\theta < 60°$. The events are selected on the basis of the topology and time compatibility of the triggered detectors. The station with the highest signal must be enclosed within an *active hexagon* in which all six surrounding detectors were operational at the time of the event. The value of $\rho$ at ground is deduced from P and T measured at the meteorological stations located at the central part of the array and at each FD site. Rather than using the raw number of triggering events, we compute the rate every hour normalized to the sensitive area, which is taken as the sum of the total area covered by the active hexagons every second. The modulation of the rate during the year, and as a function of the hour of the day, follows the changes in $\rho$ and P as shown in Fig. 1. Assuming that the rates of events computed each hour follow a Poisson distribution, a maximum likelihood fit gives the estimated values of the coefficients in eq. (3) averaged over the event distribution in the $\theta$ range $[0°, 60°]$:

$$a_P = (-0.0030 \pm 0.0003)\ \mathrm{hPa}^{-1}$$
$$a_\rho = (-1.93 \pm 0.04)\ \mathrm{kg}^{-1}\,\mathrm{m}^3$$
$$b_\rho = (-0.55 \pm 0.04)\ \mathrm{kg}^{-1}\,\mathrm{m}^3$$

corresponding to a reduced $\chi^2$ of 1.08. The result of the fit reproduces very well the daily averaged and the shorter term modulations of the measured rate of events as shown in Fig. 1.

## IV. Comparison among model, data and simulations

To complete the study of atmospheric effects, we performed full EAS simulations in different realistic atmospheric conditions. Proton-initiated EAS have been simulated at four fixed energies ($\log(E/\mathrm{eV}) = [18, 18.5, 19, 19.5]$), at seven fixed $\theta \in [0°, 60°]$ and for five atmospheric profiles (see Fig. 2), which are a parametrisation of the seasonal averages of several radio soundings carried out at the detector site [5]. The set of simulations consists of 60 EAS for each combination of atmospheric profile, energy and $\theta$.

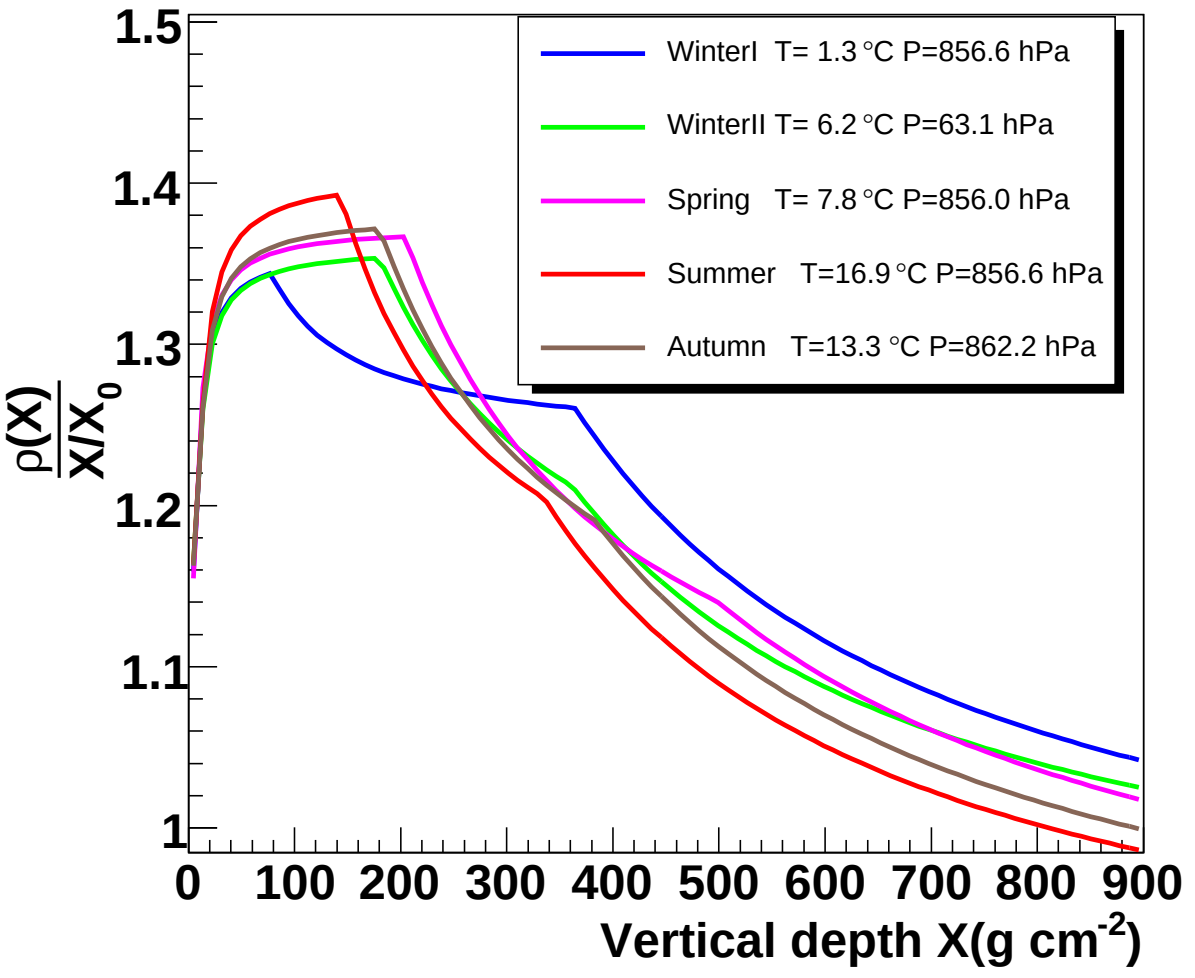

Fig. 2. Atmospheric $\rho$ profiles used in the EAS simulations normalized to an isothermal one ($X_0$ = 900 g cm$^{-2}$). These seasonal profiles come from balloon-borne sensors launched at regular intervals above the Pierre Auger Observatory site. The corresponding values of P and T are given in the box.

The comparison of the atmospheric coefficients obtained from data with those expected from the model and simulations is shown in Fig. 3. Since we are using seasonal atmospheric profiles, we do not have access to the diurnal variation of T with the EAS simulations and thus we cannot determine the $\beta_\rho$ coefficient. In the case of the data, the dependence on $\theta$ is obtained by dividing the data set in subsets of equal width in $\sec\theta$. For each subset the same fitting procedure as presented previously is used. The signal coefficients are then derived dividing the rate coefficients by $B(\gamma-1)$. Since the bulk of the triggering events have $E < 10^{18}$ eV, we used the spectral index $\gamma = 3.30 \pm 0.06$ as measured with the Pierre Auger Observatory below $10^{18.65}$ eV [6].

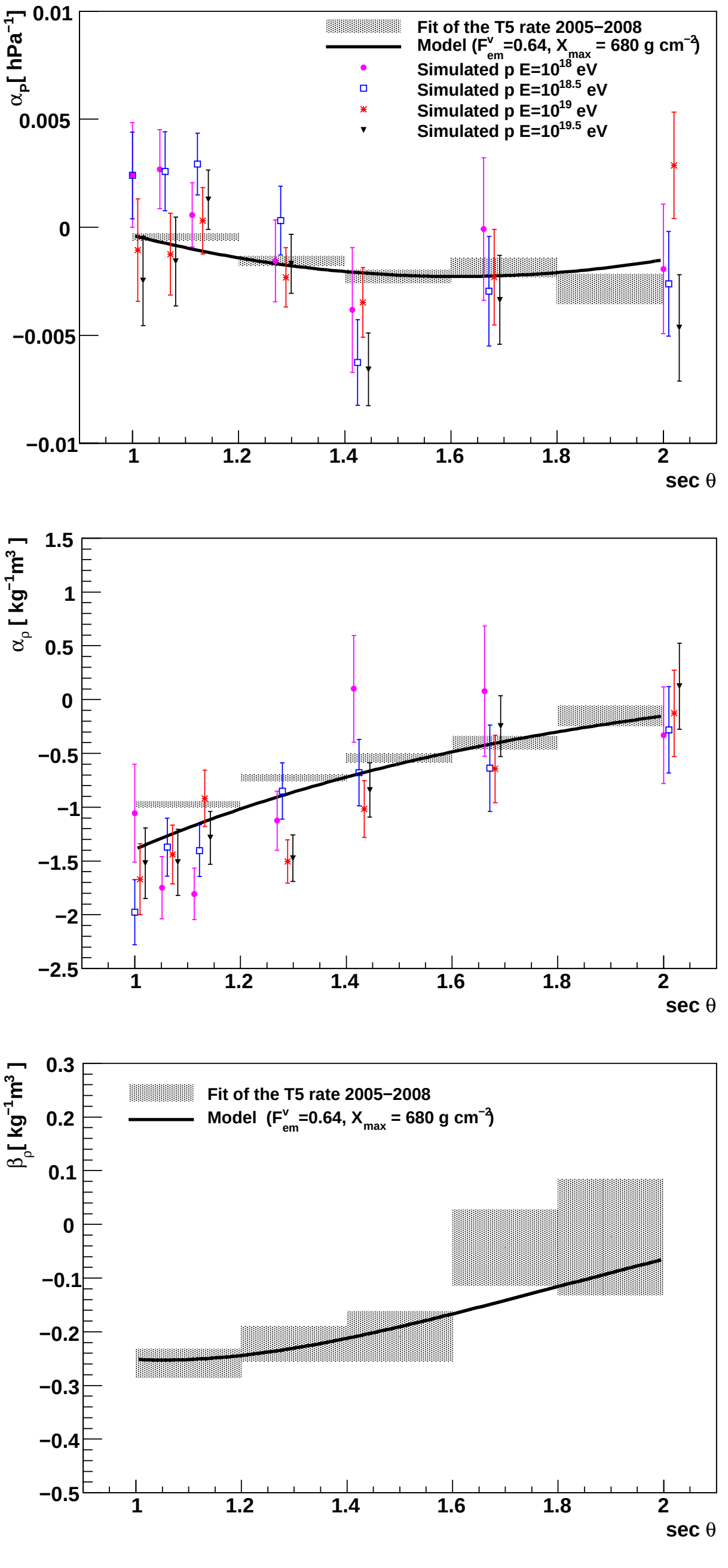

Fig. 3. Comparison of the $\alpha_P$ (top), $\alpha_\rho$ (middle) and the $\beta_\rho$ (bottom) coefficients as a function of $\sec\theta$ obtained from data (grey shaded rectangle), simulations (bullets) and model (continuous line).

## V. Correction for atmospheric effects

As explained in section II, the observed modulation in the rate of events (see Fig. 1) is due to the fact that the observed S(1000), which is used to estimate the primary energy, depends on P and $\rho$. Therefore, by applying to each event a correction of the signal, and thus of the energy, accordingly to the studied atmospheric effects, we expect to be able to obtain a non-modulated rate of events. Starting from the definition of the rate of events per unit time in a given $\theta$ bin and above a given energy:

$$R(E > E_t) = \int_{E_t}^{\infty} \mathrm{d}E \, A(E) \, \frac{\mathrm{d}J}{\mathrm{d}E}$$

the relative change in the rate of events above a given energy under changes in the atmosphere is:

$$\begin{aligned} \frac{1}{R}\frac{\mathrm{d}}{\mathrm{d}\xi} R(E > E_t) &= \frac{1}{R}\int_{E_t}^{\infty} \mathrm{d}E \, \frac{\mathrm{d}A}{\mathrm{d}\xi}\frac{\mathrm{d}J}{\mathrm{d}E} \\ &= \frac{\alpha_\xi}{R}\int_{E_t}^{\infty} \mathrm{d}E \, \frac{\mathrm{d}\epsilon}{\mathrm{d}E} \, E \, \frac{\mathrm{d}J}{\mathrm{d}E} \end{aligned}$$

where we took for simplicity $E \propto S$. Integrating by parts, we obtain:

$$\frac{\mathrm{d}R(E > E_t)}{R \, \mathrm{d}\xi} \simeq (\gamma - 1)\alpha_\xi \left( 1 - \frac{\epsilon(E_t)\int_{E_t}^{\infty} \mathrm{d}E \, E^{-\gamma}}{\int_{E_t}^{\infty} \mathrm{d}E \, \epsilon(E) \, E^{-\gamma}} \right)$$

The expression in parentheses is the relative modulation between the rate of events above a given corrected energy and the rate of events above the corresponding uncorrected signal size. We can see that, once the energy correction is implemented, no modulation in $R(E > E_t)$ is expected above the acceptance saturation[1] since $\epsilon(E) = 1$. But, in the regime where the acceptance is not saturated the acceptance of the SD for a given corrected energy will depend on P and $\rho$. This is due to the fact that when $\epsilon(E) < 1$, the energy correction is not enough anymore to correct the rate, since, depending on atmospheric conditions, the array will trigger or not: events that do not trigger the array cannot obviously be recovered.

We have implemented the energy correction on the data set described in section III. It is done on an event-by-event basis following eq. (2). The rate of events can then be computed every hour above any given corrected energy threshold. In particular, we show in Fig. 4 the rate of events during the years and as a function of the hour of the day for corrected energies greater than $10^{18}$ eV. Even if the acceptance is not saturated at $10^{18}$ eV, the trigger efficiency is still high enough and the energy correction accounts for most of the atmospheric induced systematics. Assuming Poisson fluctuations in each bin, a fit to a constant gives a reduced $\chi^2$ of 1.30 and 1.18 for respectively the seasonal and the daily rate of events that are shown in Fig. 4.

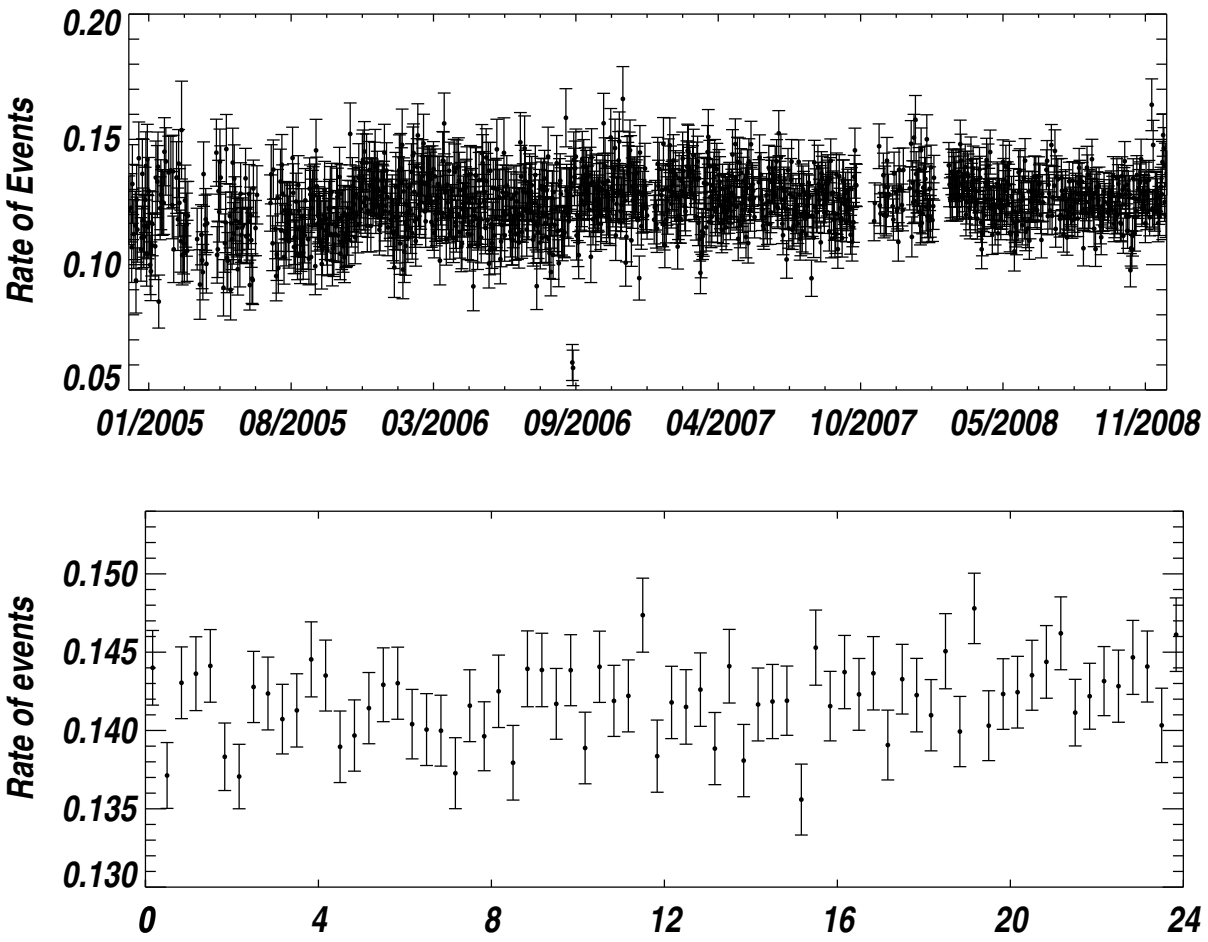

Fig. 4. Rate of events obtained above $10^{18}$ eV once the P and $\rho$ dependent conversion from signal to energy is implemented. Left: daily averaged rate of events. Right: rate of events during the day (UTC).

## VI. Conclusion

We have studied the effect of atmospheric variations on EAS measured by the array of surface detectors of the Pierre Auger Observatory. We observe a significant modulation of the rate of events with the atmospheric variables, both on seasonal scale (10%) and on a shorter time scale (2% during the day). This modulation can be explained as due to the impact of P and $\rho$ changes on the EAS development, which affect the energy estimator S(1000). Comparing the coefficients obtained from data, EAS simulations and expectations from the model built, a good agreement is reached, not only for the overall size of the effect but also for the $\theta$ dependence. By taking into account the atmospheric effects on the signal and energy estimation on a event-by-event basis, we are able to correct the observed rate of events for the seasonal modulation, thus allowing the search for large scale anisotropies at the percent level down to energies around $10^{18}$ eV [7].

[1]The SD trigger condition, based on a 3-station coincidence, makes the array fully efficient above about $3 \times 10^{18}$ eV.

# Nightly Relative Calibration of the Fluorescence Detector of the Pierre Auger Observatory

**Rossella Caruso * for the Pierre Auger Collaboration †**

*Dipartimento di Fisica e Astronomia, Università di Catania and INFN Sez.Catania, 64, Via S.Sofia, 95123-Catania,Italy
†Observatorio Pierre Auger, Av.San Martń (N)304, Malargüe, Mendoza, Argentina

***Abstract*. A relative calibration of the photomultipliers in the fluorescence telescopes at the Pierre Auger Observatory is made every night. The calibration allows the long term performance of the photomultipliers to be monitored and permits a relative calibration database to be created each night. Infrequent absolute calibrations are also performed to determine the conversion factor of photon yield to ADC counts. A stable procedure has been developed to produce absolute calibration constants, typically $2 \times 10^6$ calibration constants/year, based on the absolute calibrations but rescaled depending on the photomultiplier response on a nightly basis. Three years ($2006-2009$) of data were analysed to produce the latest version of the database, including for the first time calibration constants for the final six telescopes that were commissioned in February** 2007.

## I. The Pierre Auger Observatory

Primary particles with ultra high energy from $10^{18}$ eV to the extreme region of the GZK cutoff [1] interact in the atmosphere at nearly speed of light and create extensive air showers (EAS). The Pierre Auger Observatory [2] measures their flux, arrival direction distribution and mass composition by detecting EAS with high statistics. The Observatory includes two sites: Auger South site fully completed and operational in the Pampa Amarilla (Argentina) and Auger North site planned to be installed in Colorado (USA). The Auger South detector consists of the Fluorescence Detector (FD), 24 fluorescence telescopes collected in 4 sites on the top of natural hills, ovelooking the Surface Detector (SD), 1660 water-Cherenkov detectors deployed on a triangular grid of $1.5\,\mathrm{km}$ spacing over a wide area ($3000\,\mathrm{km}^2$). A single telescope is composed by an aperture system, a spherical mirror and a camera of 440 photomultipliers (PMTs). The signals from the PMTs are amplified, filtered and continuosly digitised by 10 Mhz 12 bit FADCs [3]. The fluorescence telescopes take data every month for a period (FD shift) of approximately 15 days, since one week before to one week after the new moon.

## II. Aim of the FD Calibration

The amount of scintillation light produced by an EAS is directly proportional to the energy deposited by the shower in the atmosphere. One FADC bin from the j$^{th}$ PMT represents light from a particular segment of atmospheric depth $\Delta X$. The conversion from energy deposited to the FADC count is given by

$$n_{ADC_j} = \frac{dE}{dX} \cdot Y_\gamma \cdot \Delta X \cdot T \cdot \frac{A}{4\pi r^2} \cdot C_j^{abs} \quad (1)$$

where $dE/dX$ is the rate of energy deposit in that segment of shower track, $Y_\gamma$ is the fluorescence photon yield per unit of energy deposit, $T$ is the atmospheric attenuation factor (mainly due to Raylegh and Mie scattering), $A$ is the telescope aperture and $r$ is the light path in the atmosphere from the EAS towards the telescope, $C_j^{abs}$ is the absolute calibration factor.
$C_j^{abs}$ depends on the optical efficiency of the telescope, on the quantum efficiency, the photoelectron collection efficiency and the gain of the PMTs and on the charge-to-digital conversion in the FADCs. An absolute and relative optical calibration of all telescopes is needed to determine the absolute convertion of photon yield to ADC counts.

## III. Overview of the FD optical calibration system

Different methods are adopted to calibrate the FD [4], among these are the absolute calibration, performed occasionally to follow the long-term behaviour, and the relative calibration performed daily to follow the short-term behaviour of the photomultiplier. The absolute end-to-end calibration [5],[6],[7] uses a cylinder with a diameter of $2.5\,\mathrm{m}$, called *drum* creating uniform illumination from an LED light source at $375\,\mathrm{nm}$. The absolute calibration of the drum is based on a $Si$ photodiode calibrated at NIST [8]. The *drum* can be mounted at each telescope entrance aperture once or twice in a year. This measurement gives the $C_j^{abs}$ conversion factor from photons to ADC counts (eq.1).
Three different (**Cal A**, **Cal B**, **Cal C**) relative optical calibrations [9] are performed to monitor different parts of the telescope, its daily performance and time variations between two subsequent absolute calibrations.
• In the **Cal A** calibration (fig.1), the light pulses are produced with a bright ($470\,\mathrm{nm}$) LED, transmitted from the source to a $1\,\mathrm{mm}$ thick Teflon diffuser located in the center of the mirror. The light illuminates directly the camera. This calibration monitors only the behaviour of the photomultipliers.

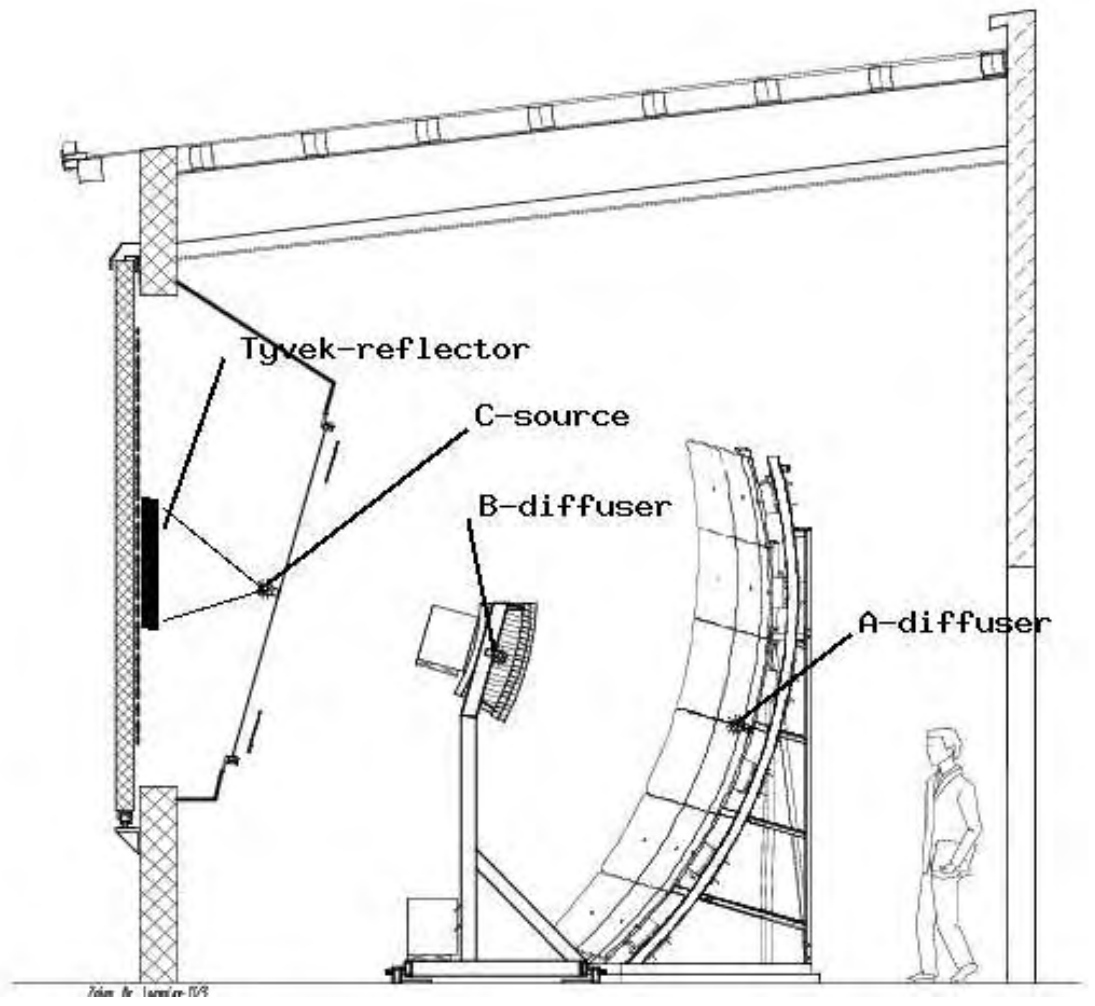

Fig. 1: A schematic lay-out of the fluorescence telescope of the Pierre Auger Observatory. The location of the light diffusers corresponding to the three **Cal A**, **Cal B**, **Cal C** optical relative calibrations is shown.

• In the **Cal B** calibration (fig.1), the light pulses are produced with a Xenon flash lamp and transmitted to a 1 mm thick Teflon diffusers located on the center of two sides of the camera. The light illuminates the mirror and then is reflected to the camera. This calibration is aimed at checking the change in the reflectivity of the mirror and the behaviour of the PMTs.

• In the **Cal C** calibration (fig.1), the light pulses are flashs from a Xenon lamp to diffusers located just outside the entrance aperture. The light illuminates a reflective, removable Tyvek screen inserted outside the UV filter and then is reflected back towards the mirror. This calibration is intended to check the whole chain through the filter, reflection by the mirror and the behaviour of the PMTs.

The relative calibration measurements are performed twice per night, at the beginning and at the end of the FD data acquisition, to track variations throughout the data taking, for every night during the FD shift.

## IV. Nightly Cal A Relative Calibration

To perform one **Cal A** calibration measurement, 50 LED pulses ($N_{LED}$) at a rate of $1/3$ Hz with square-type waveform (fig. 2) are generated. The FD data acquisition is triggered externally and the PMT signals are stored in files of 25 MB size.

For a given telescope, calibration raw data are processed to extract the mean integral charge $< Q^{CalA} >_{j,k}$ for the $j^{th}$ photomultiplier computed as the average over $N_{LED}$ in the k$^{th}$ calibration measurement:

$$< Q^{CalA} >_{j,k}= \sum_{i}^{N_{LED}=50} Q^{CalA}_{i,j,k}/N_{LED} \; ; \qquad (2)$$

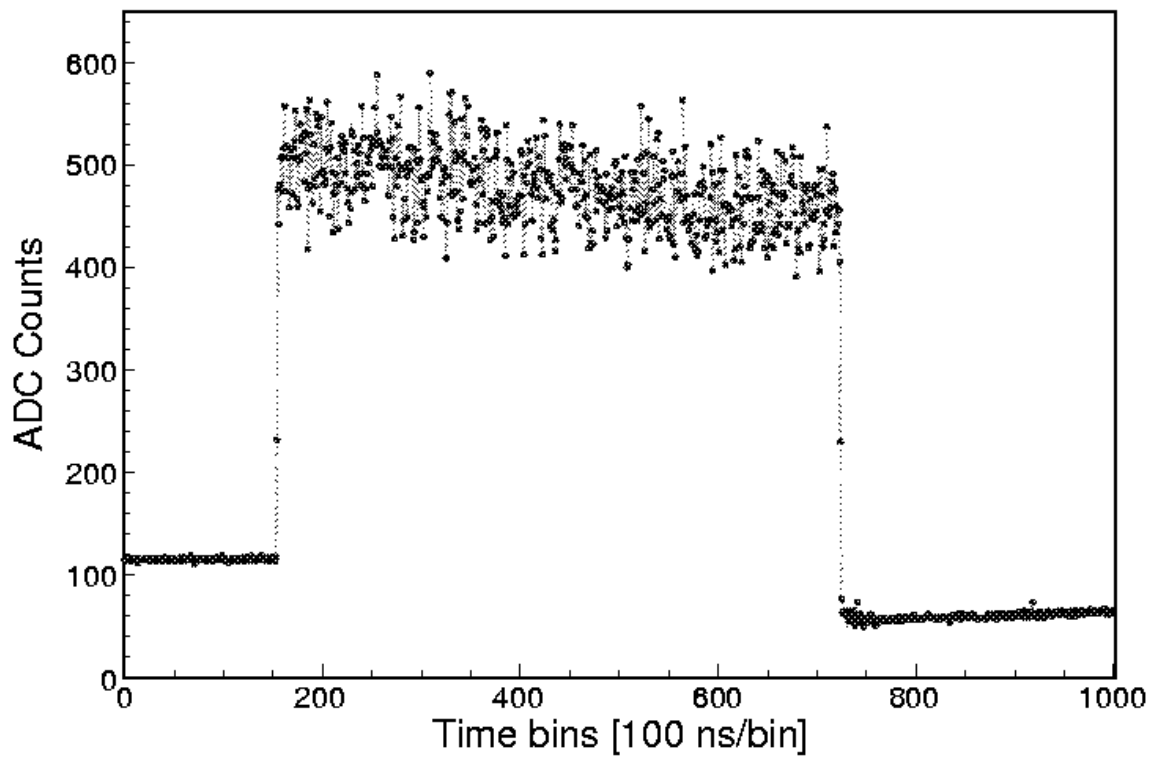

Fig. 2: The **Cal A** light pulses have a square wave form with a typical width of $57\,\mu$s and a drop in the height introduced by the control loop for very long-lasting signals. The difference in the pedestal before and after the pulse is due an undershoot of the system.

where

$$Q^{CalA}_{i,j,k} = \left( \sum_{l=t_{start}}^{l=t_{stop}} n_{ADC_l} \right)_{i,j,k} \qquad (3)$$

is the sum of $n_{ADC}$ FADC counts for the $i^{th}$ LED pulse, subtracted the signal pedestal, in the $t_{start} \leq l \leq t_{stop}$ integration gate; $l$ is a single $100\,ns$ FADC time bin , $t_{start}$ and $t_{stop}$ are respectively the first and the last $l$ where the signal can be considered over threshold according to given conditions. Different algorithms to scan the **Cal A** signal have been developed. Consistency cross-checks have been carried out and have shown an excellent agreement among different codes.

## V. The relative calibration constants

The **Cal A** data amount (about $40$ GB equal to $1.5 \times 10^3$ files per year per telescope) is stored partly on tapes and partly on disks at the Pierre Auger Observatory. In order to reduce the impact of data transfer on the existing

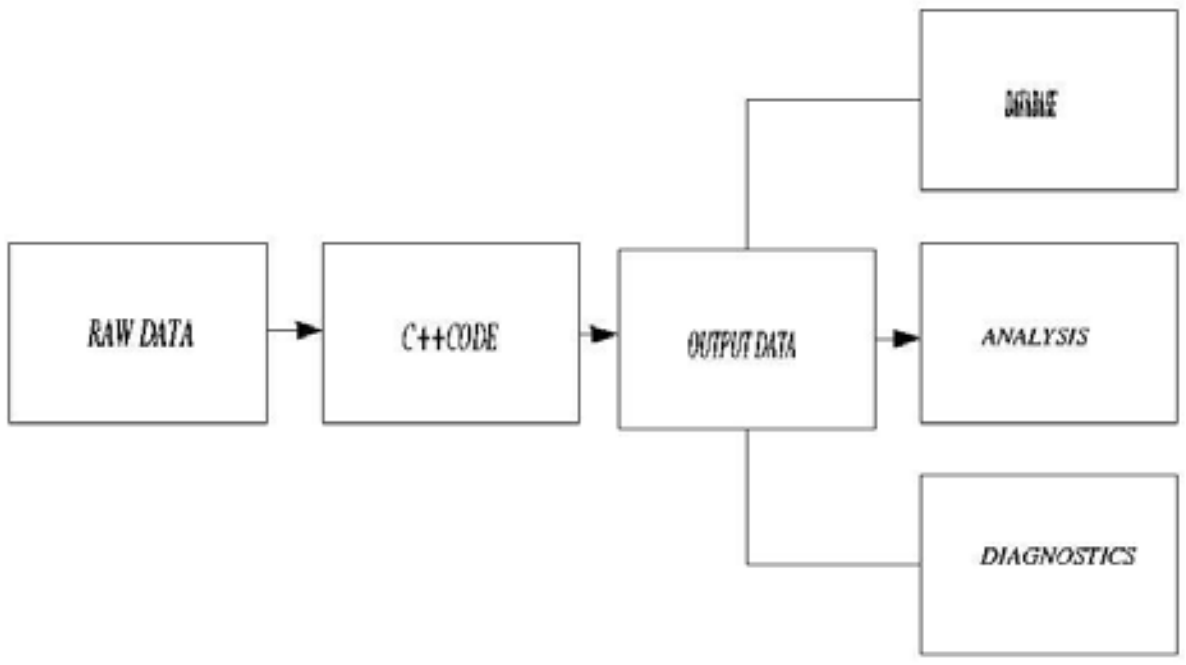

Fig. 3: Block scheme of the data processing and the further off-line analysis.

Internet link, the **Cal A** data processing is performed in the Pierre Auger Observatory and only the output results transferred outside of the Observatory for the further off-line analysis (fig. 3). For each **Cal A** calibration measurement, the value in the eq.3 is normalized by using the same quantity $< Q_j^{CalA} >_{ref}$ calculated for a given reference run:

$$C_{j,k}^{rel} = \frac{< Q^{CalA} >_{j,k}}{< Q^{CalA} >_j^{ref}} \quad (4)$$

The reference run (one per each telescope) is taken within one hour after the absolute calibration measurement. The ratio in eq.4 is the $C_{j,k}^{rel}$ *relative calibration constant* for the k$^{th}$ calibration measurement. It represents the relative change in absolute gain of the j$^{th}$ photomultiplier. The relative calibration constant fluctuates around the nominal value (equal to 1) with a typical r.m.s. of a few percent.

## VI. Monitoring the stability of the Fluorescence Detector

The **Cal A** relative calibration allows the short- and long-term behaviour of the photomultipliers to be monitored. Three years ($\mathrm{March\,2006} - \mathrm{March\,2009}$) of data have been analysed for all the telescopes. The overall stability of the 24 fluorescence telescopes has been carefully and systematically studied.

The telescopes are quite stable on short-term, showing a $2-3\,\%$ variation within each night (fig. 4a) and a $1-3\,\%$ variation within each FD shift (fig. 4b), apparently induced by night sky exposure. On medium- and long-term, since the beginning of year 2007, owing to more restricting prescriptions in operation conditions, the FD response appears stable. The overall uncertainty, as deduced from the medium-term (approximately six months) monitoring, is typically in the range of $1-3\,\%$ (fig.5). In addition, seasonal variations of $3-4\,\%$ have been observed in all telescopes, likely due to temperature variations in the buildings lodging the telescopes (fig.5). The observed loss of gain, averaged over all telescopes, is less than $2\,\%$ per year. It does not affect the life time of the FD so far. The system is currently very stable.

## VII. Production of the nightly absolute data base

The **Cal A** relative calibration permits an absolute calibration database (DB) to be created each night. To compensate for the short- and long-term variations in the telescope response and to minimize calibration uncertainties, absolute calibration constants $C_{j,k}^{abs^\star}$ for the j$^{th}$ photomultiplier and the k$^{th}$ calibration measurement are produced on a nightly basis. They are based on the absolute calibrations but are rescaled depending on the PMT response according to

$$C_{j,k}^{abs^\star} = \frac{C_j^{abs}}{C_{j,k}^{rel}} = \frac{< Q^{CalA} >_{j,k}^{ref}}{< Q^{CalA} >_{j,k}} \cdot C_j^{abs} \quad (5)$$

To produce the nightly DB, only **Cal A** relative calibration measurements acquired at the end of the FD data taking are selected as their response is more stable.

A steady procedure has been developed to produce nightly absolute calibration constants. Three years ($2006 \div 2008$) of calibration data for all the telescopes, including for the first time the final six telescopes that were commissioned in February 2007, have been used to produce the latest version of the nightly database. It contains about $6 \times 10^6$ absolute calibraticn constants, its size is $0.9\,\mathrm{GB}$. In the current DB, a flag is assigned to each PMT to record the goodness of the corresponding calibration constant, according to criteria that take any hardware or software failure in the calibration system, in the camera or in the front-end electronics into accont. The $99.5\%$ of calibration constants comes out to have an expected value, only the $0.5\%$ of them is out of range and has to be rejected. Lastly, before each release of new costants, physics tests are performed and their outcome compared with known references to validate them.

## VIII. Conclusions

Three years of relative calibration measurements for all the fluorescence telescopes of the Pierre Auger Observatory have been analysed. The short- and long-term behaviour of the photomultipliers of the Fluorescence Detector has been monitored. A steady procedure has been developed to produce $6 \times 10^6$ nightly absolute calibration constants, including for the first time the final six telescopes commissioned in February 2007.

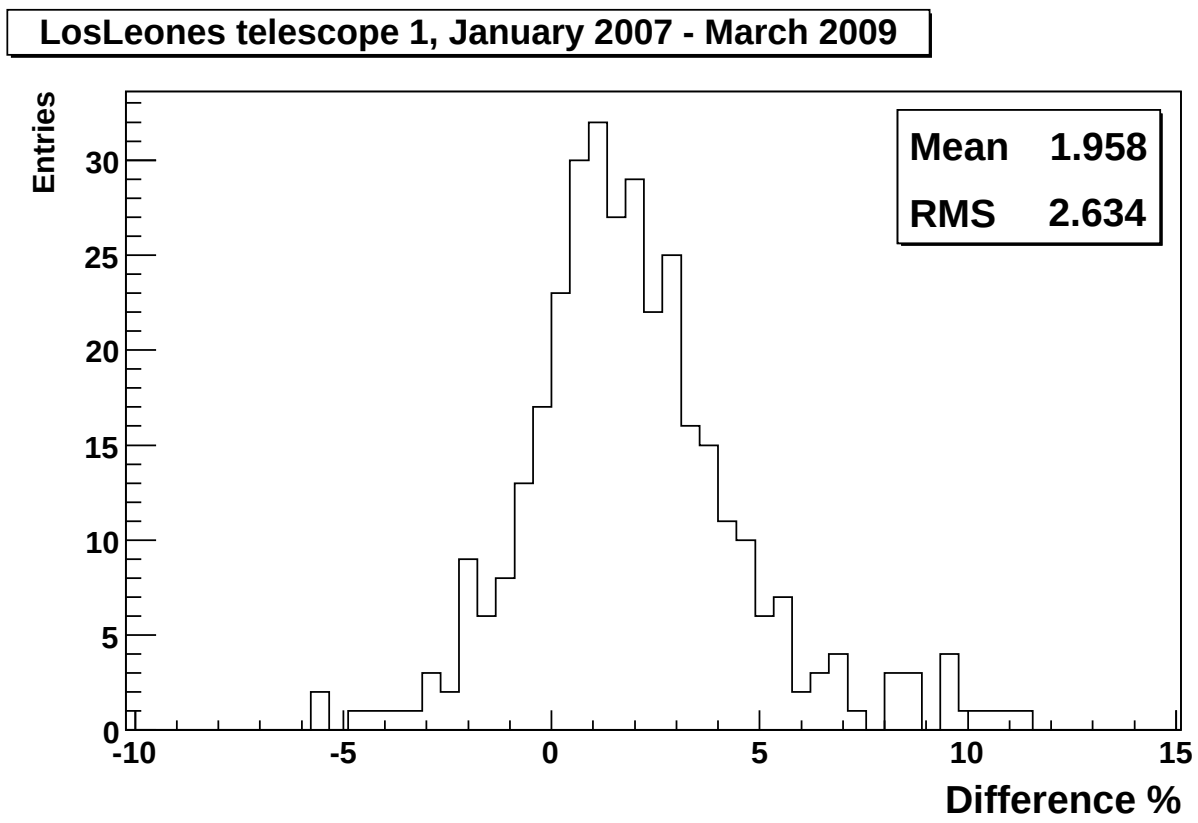

[a)]

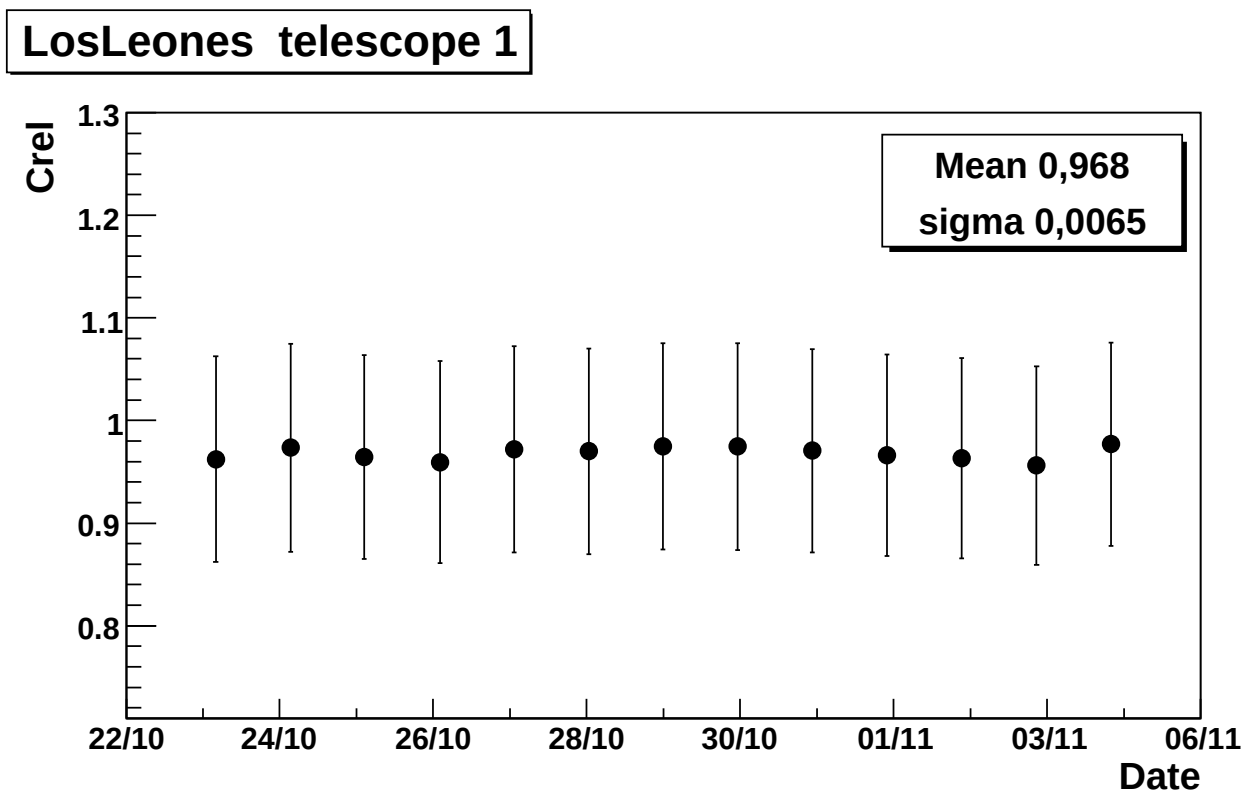

[b)]

Fig. 4: Variations on the short-time behaviour of the photomultipliers within each night [a)] and each FD shift [b)]. On the left, the distribution shown is based off the difference per cent between relative calibration measurements acquired at the beginning and at the end of a given FD nightly data taking. On the right, the results are shown for a period of approximately 15 days (FD shift) of data taking. For each night in the FD shift, they are obtained by averaging over all 440 photomultipliers in the telescope. The error bars represent the one sigma fluctuations of the relative calibration constants over all the 440 PMTs.

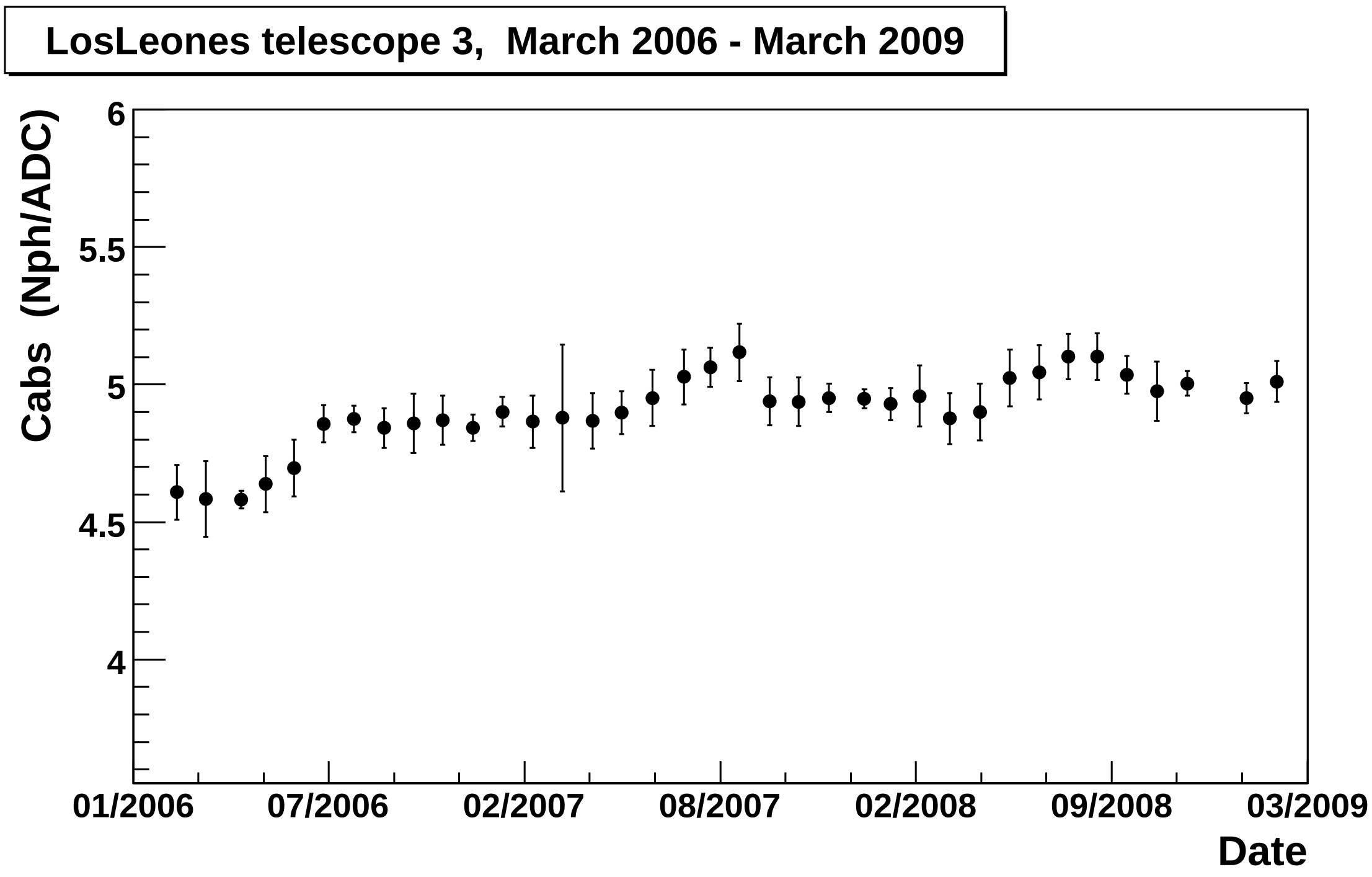

Fig. 5: Variations on the long-time behaviour of the photomultipliers over three years of the FD data taking. The results show the nightly absolute calibration constants as a function of the time. Each point is the mean value obtained by averaging over all the 440 photomultipliers in the telescope and over all the nights during one FD shift. The error bars represent the one sigma night to night fluctuations during each FD shift.

# Rapid atmospheric monitoring after the detection of high-energy showers at the Pierre Auger Observatory

**Bianca Keilhauer***, **for the Pierre Auger Collaboration**†

*Karlsruhe Institute of Technology (KIT),
Forschungszentrum Karlsruhe, Institut für Kernphysik, P.O.Box 3640, 76021 Karlsruhe, Germany
†Observatorio Pierre Auger, Av. San Martin Norte 304, 5613 Malargüe, Argentina

***Abstract.*** **The atmospheric monitoring program of the Pierre Auger Observatory has been upgraded to make measurements of atmospheric conditions possible after the detection of very high-energy showers. Measurements of the optical transmittance due to aerosols and clouds are time-critical. Therefore, observations of atmospheric regions close to a shower track of interest are performed within ten minutes of a shower detection using LIDAR and telescope monitors. Measurements of the altitude dependence of atmospheric state variables such as air temperature, pressure, and humidity are performed within about two hours following the detection of a very high-energy event using meteorological radio soundings. Both programs are triggered using a full online reconstruction with analysis-level quality cuts. We describe the implementation of the online trigger, and discuss the impact of the monitoring data with high resolution on the analysis of air shower events.**

## I. Introduction

At the Pierre Auger Observatory [1], extensive air showers (EAS) induced by ultra-high energy cosmic rays are studied. The observatory consists of two detector types, a surface detector (SD) for secondary particles of EAS and fluorescence detector (FD) telescopes for UV-emissions by nitrogen molecules in the atmosphere. The fluorescence technique provides an almost calorimetric measurement of the primary energy of cosmic rays.

However, the constantly changing conditions of the atmosphere demand a sophisticated monitoring system [2]. The reconstruction of air showers from their UV-emission requires proper characterisation of atmospheric state variables such as pressure, temperature, and humidity, as well as the optical transmittance due to aerosol contamination and the presence of clouds [3]. The state variables of the atmosphere above the Pierre Auger Observatory are determined using meteorological radio soundings, while aerosol and cloud conditions are measured by two central lasers, four elastic LIDARs, and four cloud cameras [4].

The sounding data have been incorporated into monthly models, and aerosol and cloud data into an hourly database [4]. However, for events of particular physical interest, such as very high-energy showers, it is desirable to measure the properties of the atmosphere as accurately as possible. To improve the resolution of the atmospheric database for such events, dedicated radio soundings and LIDAR measurements can be triggered by an online event reconstruction. We will discuss the motivation for such measurements (Section II), the operation of the online trigger (Section III), and the use of dedicated atmospheric measurements in the offline reconstruction (Section IV).

## II. Motivation for Rapid Monitoring

Between 2002 and 2005, radio soundings were performed at the observatory during dedicated measurement campaigns. Since mid-2005, the soundings have been performed approximately every fifth day. The measurements obtained by launching weather balloons provide altitude profiles of the air temperature, pressure, and relative humidity up to about 23 km above sea level. Due to the limited statistics of the measurements, the data have been incorporated into monthly models of conditions near Malargüe, Argentina, the site of the southern part of the Pierre Auger Observatory [4], [5].

Using monthly models instead of actual profiles introduces an uncertainty of the primary energy of $\Delta E/E = 1.5\% - 3\%$ for showers with energies between $\approx 10^{17.7}$ eV and $10^{20}$ eV, and a corresponding uncertainty $\Delta X_{\mathrm{max}} = 7.2 - 8.4$ g cm$^{-2}$ of the position of the shower maximum. While it is not practical to perform a radio sounding every night, the reconstruction can be improved for a subset of the EAS data by concentrating the soundings in periods when high-quality events are observed. This subset of EAS events is particularly important because they contribute to the energy scale determination of the entire observatory [6].

For aerosol measurements, the LIDAR stations conduct automated hourly sweeps of the atmosphere above the observatory to estimate the vertical aerosol optical depth, cloud height, and cloud coverage [7]. The hourly sweeps are sufficient to characterise changing aerosol conditions, but a more rapid response is necessary to identify moving clouds between shower tracks and the FD telescopes observing the event. To accomplish this, the LIDARs are capable of interrupting their hourly sweeps to scan interesting shower tracks for atmospheric non-uniformities [7], [8].

## III. Online Trigger

To select events for monitoring with radio soundings and/or LIDAR scans, an online reconstruction is used to trigger balloon launches and the LIDAR hardware. As data are acquired from the FD telescopes and SD, they are merged by an event builder into hybrid event files, and passed to the reconstruction software. The software is the same as that used for Off$\underline{\text{line}}$ event reconstruction [9], including the latest versions of the detector calibration databases. In this way, the LIDAR and balloon triggers can be constructed with the same quality as the offline physics analysis.

The reconstruction loop runs every 60–90 seconds, and reconstructs events between 2 and 10 minutes after their detection[1]. Events with reconstructible $dE/dX$ longitudinal profiles are used to trigger LIDAR and sounding measurements following the application of basic quality cuts. The LIDARs trigger on showers with $E \geq 10^{19}$ eV in combination with given quality cuts on the reconstruction of the shape of the longitudinal profile. These events are typically of high quality and the rapid monitoring is to ensure that no atmospheric impurity has altered the reconstruction result. To allow the investigation of shower observations affected by clouds and other non-uniformities in the atmosphere for possible longitudinal profile corrections in the future, few events of lower quality with $E \gtrsim 10^{18.78}$ eV can also pass the trigger conditions. This yields up to one scan per night. A balloon launch is triggered for events with $E \geq 10^{19.3}$ eV and a profile fit $\chi^2$/NDF $< 2.5$. All trigger conditions have in common that the position of shower maximum has to be well in the field of view and that the observed track has an expedient length.

The quality of the online reconstruction has been checked by comparing with results from the Off$\underline{\text{line}}$ reconstruction. Even though some minor differences in the reconstruction chains are present, the reconstruction quality is excellent. Only some events are missed by the online reconstruction below $10^{18}$ eV, which is well below the required energy threshold for both rapid monitoring programmes. At primary energies of interest, the energy of the primary cosmic ray and the position of the shower maximum are reconstructed very well by comparison with the Off$\underline{\text{line}}$ reconstruction: only below 1% difference for the energy and 2 g cm$^{-2}$ in $X_{\max}$ are expected. The reconstruction cuts for triggering radio soundings yield a trigger rate of 3 to 13 radio soundings per shift[2] depending on season, see Fig 1. In practice, only one launch is performed within 5 hours resulting in about 2 to 6 launches per FD shift.

Triggers for the LIDAR systems are handled automatically by these stations: the hourly scans are halted and the LIDARs sweep into the field of view of the FD telescopes to probe the shower track [7]. To avoid triggering the telescopes with stray light, the FD data acquisition is vetoed for four minutes, the maximum duration of a dedicated scan. In contrast to the LIDAR, the balloon launches require human intervention. Therefore, a sounding trigger initiates a SMS text message to a technician in Malargüe. The technician then drives to the balloon launching facility and performs the sounding typically within two hours of the detection of the event. This measurement has no interference with any other data acquisition of the Pierre Auger Observatory.

[1]The delay is caused by buffering of station data from the SD.

[2]To infer these numbers, the EAS data sample from 2008 was analysed.

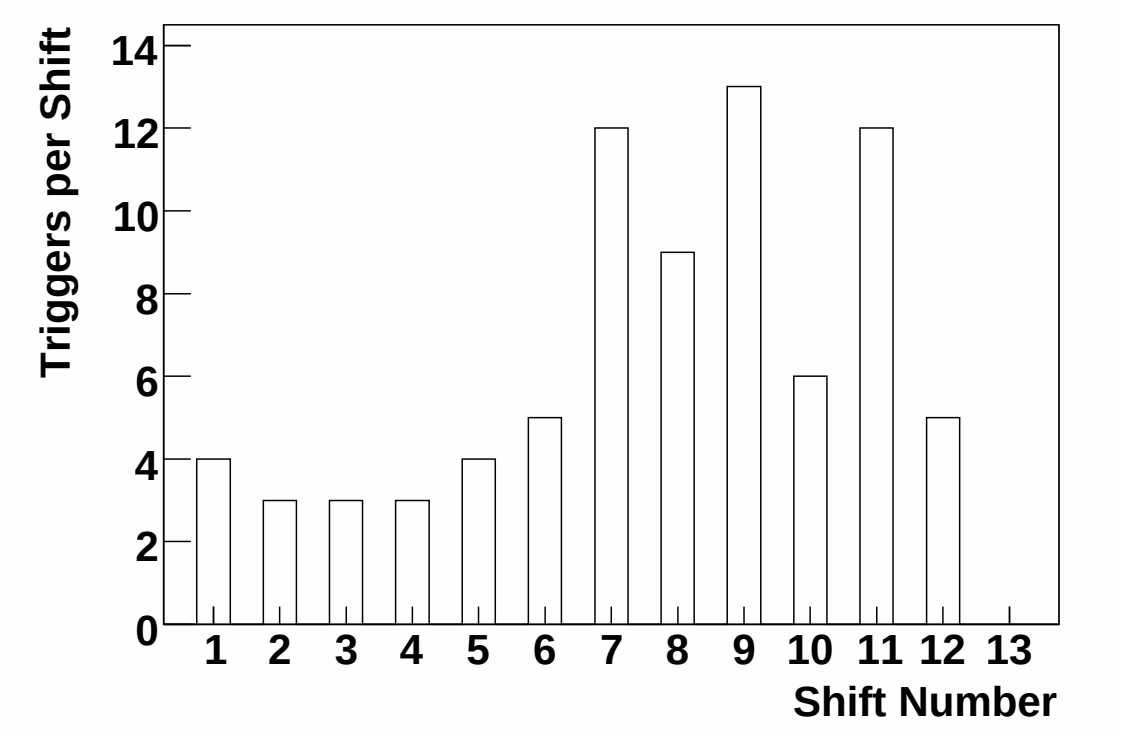

Fig. 1. All triggers for each FD shift in 2008 of events which would have passed the sounding trigger conditions. A seasonal effect due to longer nights in winter can be seen.

## IV. Analysis

During the March – April 2009 FD shift, the rapid monitoring with radio soundings was activated for the first time. We had two nights with successful triggers for the radio soundings. In the second night, it was a stereo event. Both radio soundings could be performed within 1.5 hours after the high-energy air shower. The first trigger was sent at the end of March and the second one at the beginning of April. In Fig. 2, the difference between the actual measured atmospheric profiles from the radio soundings and the monthly models for the area of the Auger Observatory valid for that month are displayed for the temperature, atmospheric depth, and vapour pressure. For the event in March, the differences between the measured temperature and atmospheric depth profiles and the monthly average model are small. However the considerable amount of water vapour in the lower atmosphere indicates possible distortions of the longitudinal shower profile compared with a reconstruction using the adequate monthly model. A reconstruction of the first event with the actual atmospheric profiles compared with that using monthly models yields a $\Delta E/E$ of +0.9% and a $\Delta X_{\max}$ of +6 g cm$^{-2}$. For the event in April, the water vapour content is nearly the same as in the corresponding monthly model. However, the higher temperature close to ground resulting in lower atmospheric depth values will change the reconstructed air shower event. The same two versions of reconstruction as for the first event yield a $\Delta E/E$ of -0.5% and -1.0% for the two different FD stations which observed this stereo event and a $\Delta X_{\max}$ of +4 g cm$^{-2}$ and +3 g cm$^{-2}$.

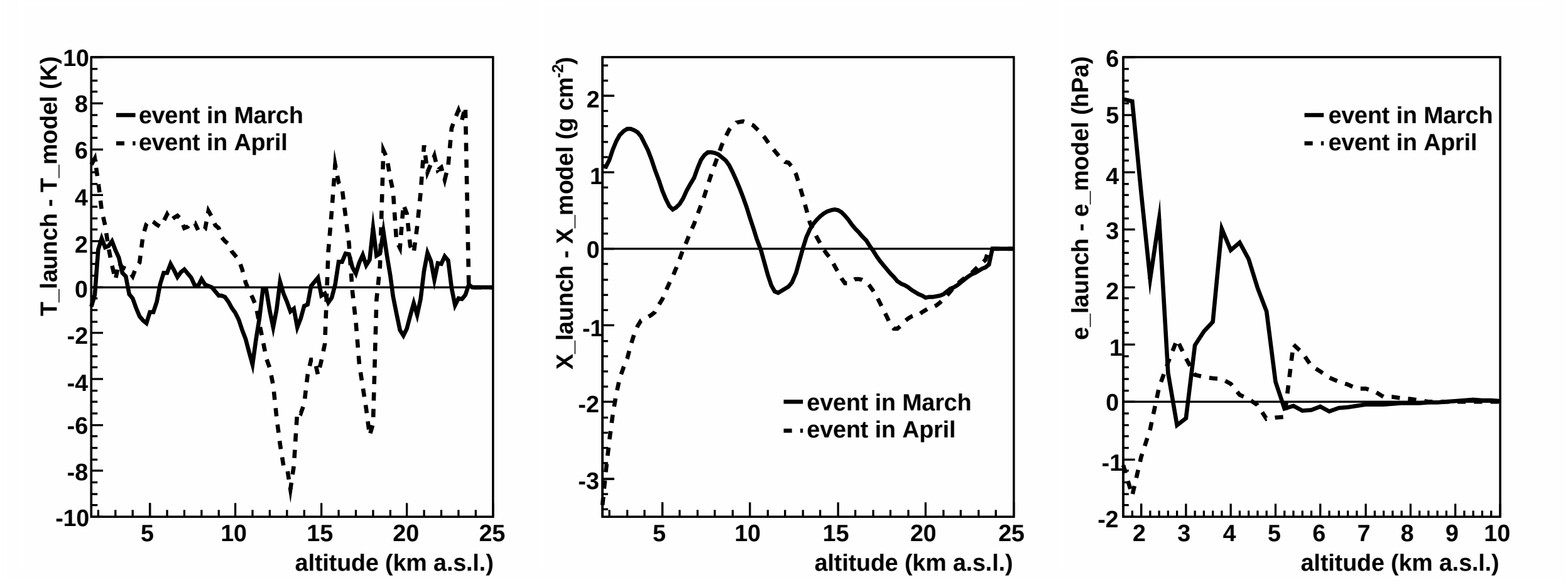

Fig. 2. Difference between two actual measured atmospheric profiles in March and April 2009 from the radio soundings and the corresponding monthly models for the area of the Auger Observatory. Left: Temperature. Middle: Atmospheric Depth. Right: Vapour Pressure.

In the second shift running this programme, we had 10 triggers in 6 nights. The first one was again a stereo event and in the fourth night, there were 3 triggers within 2.5 hours. The fifth night also provided two triggers in 2.5 hours, and in the last night there were 2 triggers within 1 hour. In total, we had 5 radio soundings initiated by high-energy air shower events, because the SMS during the last night were lost.

All events have been reconstructed using two different configurations. The first one represents the status of currently best knowledge, so using the actual atmospheric profiles from the radio soundings in combination with descriptions of fluorescence emission [10] and transmission taking into account all temperature, pressure, density, and humidity effects. The second reconstruction relies on the same descriptions but uses the monthly models for the site of the Pierre Auger Observatory which provide also profiles of water vapour. In Fig. 3, the resulting differences of the reconstruction procedures are shown for all events during March and April 2009. The stereo events have been reconstructed independently for the two FD stations which observed the extensive air shower. The primary energies of these events vary from the threshold energy up to almost $10^{19.7}$ eV and for the position of shower maximum, values between 654 and 924 g cm$^{-2}$ slant depth are observed. The given differences are between reconstruction with actual atmospheric profiles and that with monthly models. For the primary energy, we expect an uncertainty of $\pm$ 2.5% at $E_0 = 10^{19.3}$ eV while using monthly models. The differences between reconstructions using sounding data and the monthly models fit these expectations (Fig. 3 left). For the position of shower maximum, the expected uncertainty at $E_0 = 10^{19.3}$ eV is $\pm$ 8 g cm$^{-2}$. The reconstruction with monthly models nearly matches these expectation but is biased to one direction for this season (Fig. 3 right).

The rapid monitoring with LIDARs started in February 2009 and through the beginning of May 2009, the four LIDAR stations at the Pierre Auger Observatory were triggered 29 times. The intention is to investigate atmospheric conditions for those high-energy showers that fail strict analysis cuts due to distortions caused by clouds and aerosols.

For high-energy showers of high reconstruction quality, the LIDAR scans can be used to verify the quality of the atmosphere. In this manner, the scans allow for the investigation of atmospheric selection effects on the highest energy showers. Of the 29 showers probed by dedicated LIDAR scans, 17 passed the strict quality cuts used in the analysis of FD data. The energies of these showers ranges from $10^{19}$ to $10^{19.52}$ eV. The observed shower maxima are between 678 and 808 g cm$^{-2}$.

In nearly all cases, the profile fit is of high quality, and the LIDAR data do not indicate the presence of large amounts of aerosols or heavy cloud coverage. One exception is shown in Fig. 4, in which the light from the upper segment of a shower track is blocked by a thick cloud layer. The backscattered light from the LIDAR scan shows a strong echo near 8 km above ground level, or 650 g cm$^{-2}$ slant depth along the shower track, confirming the presence of a cloud.

At present, the rapid monitoring with LIDARs is mainly used as a check of the quality of the atmosphere after the observation of high-energy showers. This is quite important for analyses that rely on unusual features in shower tracks, such as exotic particle searches. The LIDAR shots can also be used to remove obscured or distorted sections of a shower track from the analysis. Once sufficient statistics have been collected, it should be possible to use the LIDAR data to correct observed shower tracks for inhomogeneities in the atmosphere.

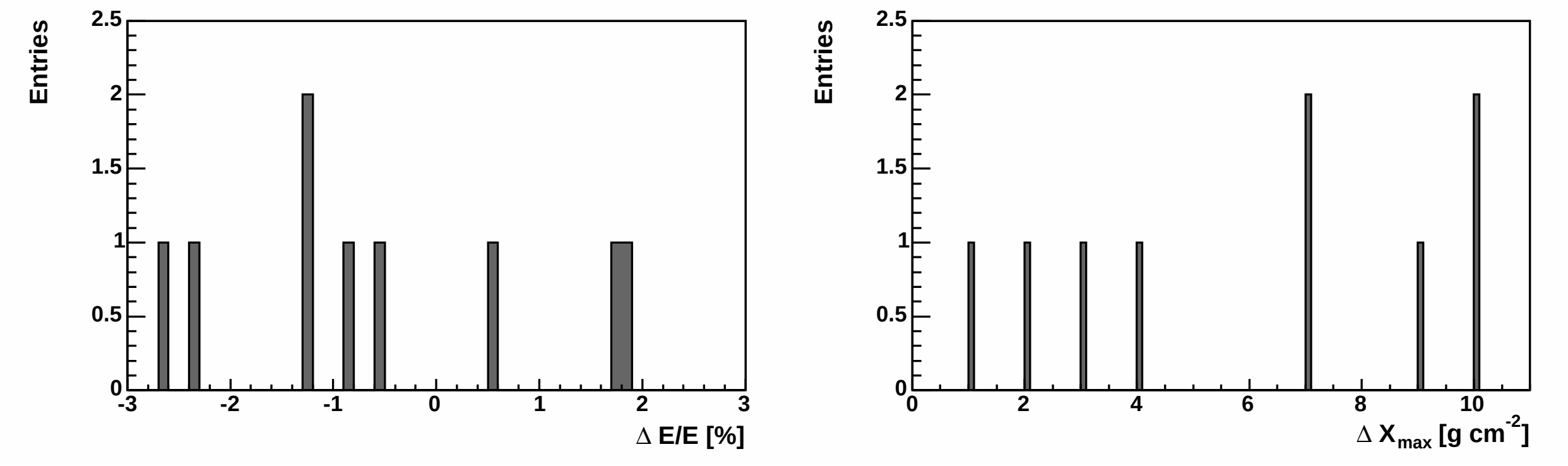

Fig. 3. Comparison of two different versions of reconstruction for air shower events observed in March and April 2009. The first reconstruction uses actual atmospheric profiles from radio soundings performed shortly after the detection of the EAS. The second one uses monthly models developed for the site of the Pierre Auger Observatory.

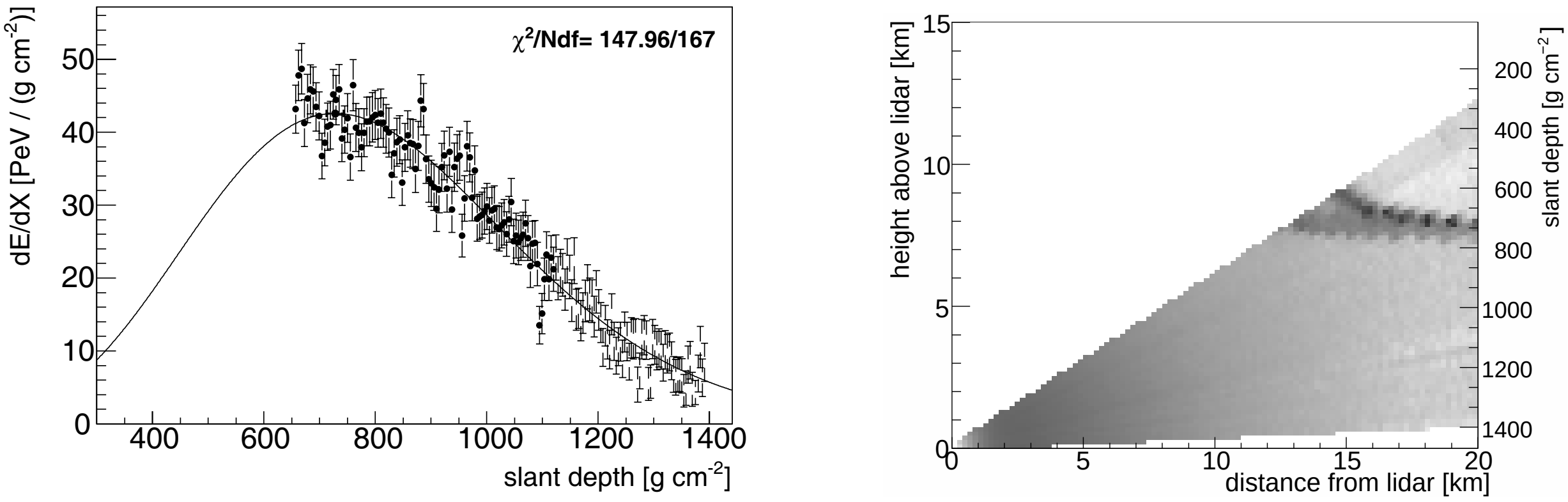

Fig. 4. Left: A $10^{19.48}$ eV shower profile obscured below 650 g $cm^{-2}$ by a cloud. The backscattered light from a LIDAR scan of the shower-detector plane (right) confirms the presence of a cloud layer (the dark horizontal band) in the telescope field of view.

# Atmospheric Aerosol Measurements at the Pierre Auger Observatory

**Laura Valore*** **for the Pierre Auger Collaboration**

*Universitá degli Studi di Napoli ”Federico II” and INFN Napoli*

***Abstract*. The Pierre Auger Observatory uses the atmosphere as a huge calorimeter. This calorimeter requires continuous monitoring, especially for the measurements made with the fluorescence telescopes. A monitoring program with several instruments has been developed. LIDARs at the sites of each of the fluorescence detectors are used to measure aerosols and clouds. Beams from calibrated laser sources located near the centre of the Observatory are used to measure the light attenuation due to aerosols, which is highly variable even on time scales of one hour. The Central Laser Facility (CLF) has been used to provide hourly aerosol characterisations over five years based on two independent but fully compatible procedures. The eXtreme Laser Facility (XLF), located in a symmetric position relative to the CLF and the four fluorescence detector sites, has just started operation. The level of cloud coverage is measured using cameras sensitive to the infrared and can also be detected with the sky background data.**

## I. Introduction

Primary cosmic rays at ultrahigh energies (E $>$ $10^{18}$eV) cannot be observed directly because of their extremely low flux. The properties of primary particles (energy, mass composition, arrival direction) are deduced from the study of the cascade of secondary particles originating from their interaction with air molecules. The Pierre Auger Observatory is a hybrid detector with an array of more than 1600 surface detectors overlooked by 24 fluorescence telescopes grouped in 4 sites each with 6 telescopes at the array periphery. The Fluorescence Detector (FD) is designed to perform a nearly calorimetric measurement of the energy of cosmic ray primaries, since the fluorescence light emitted by nitrogen air molecules excited by shower charged particles is proportional to the energy loss of the particles. Due to the constantly changing properties of the calorimeter (i.e. the atmosphere), in which the light is both produced and through which it is transmitted, a huge system with several instruments has been set up to perform a continuous monitoring of its properties. In particular aerosols are highly variable on a time scale of one hour. We perform measurements of the aerosol parameters of interest: the aerosol extinction coefficient $\alpha(h)$, the vertical aerosol optical depth $\tau_a(h)$, the normalised differential cross section $P(\theta)$ (or phase function), and the wavelength dependence of the aerosol scattering parametrised by the Ångstrom coefficient $\gamma$. Recent results showing that cloud coverage has a major influence on the reconstruction of air showers has led to a special effort in clouds monitoring.

## II. The Aerosol Monitoring System

The Pierre Auger Observatory operates an array of monitoring devices to record the atmospheric conditions. Most instruments are used to estimate the hourly aerosol transmission between the point of production of the fluorescence light and the Fluorescence Detectors and for the detection of clouds. If not properly taken into account, these dynamic conditions can bias the showers reconstruction. A map of the Pierre Auger Observatory aerosol monitoring system is shown in fig. 1.

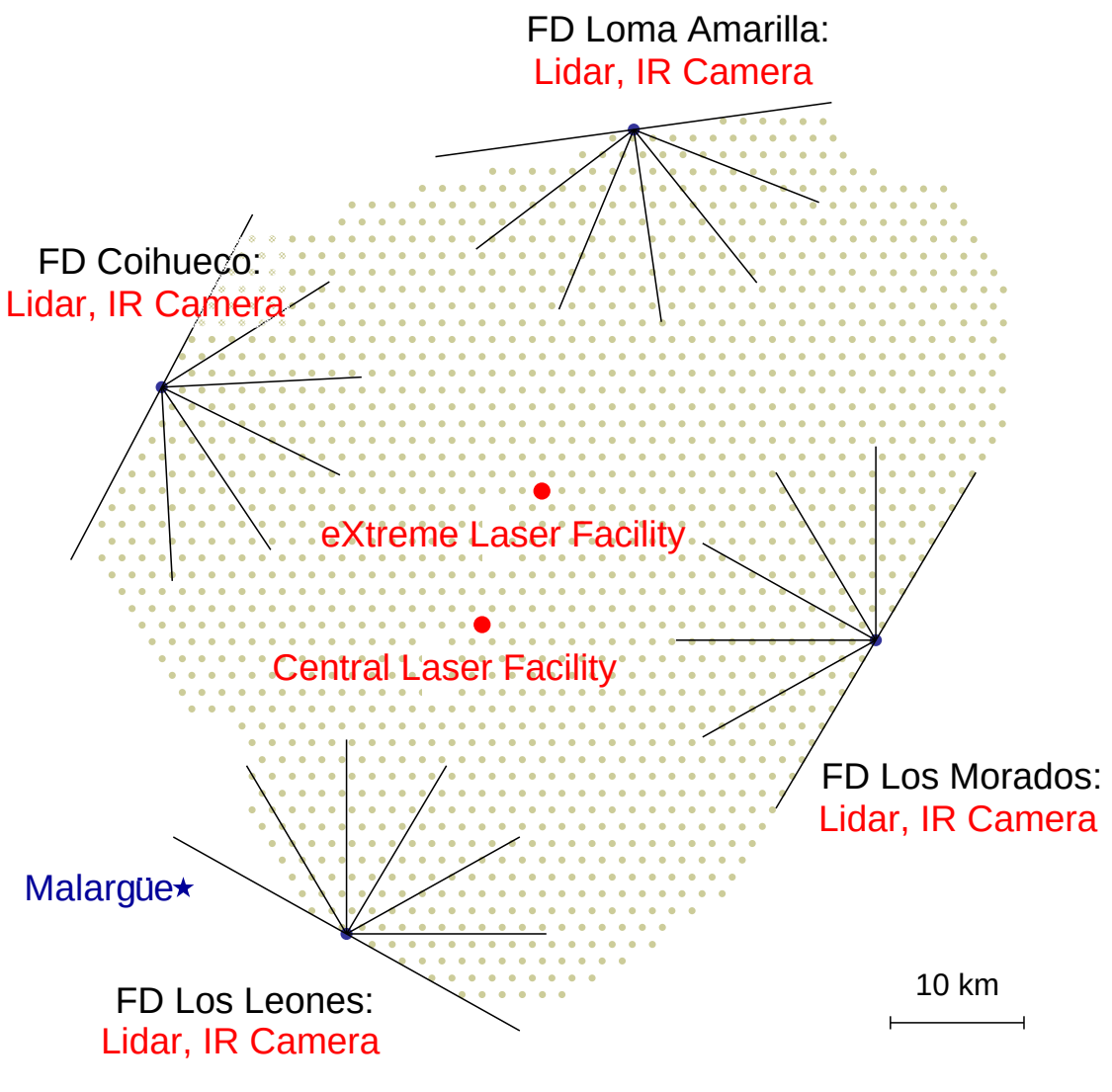

Fig. 1. Atmospheric monitoring devices map

In this paper, systems measuring aerosol optical depth and clouds, which are the main sources of uncertainties, are described. The aerosol optical depth contributes to the uncertainty on energy from 3.6% at E = $10^{17.5}$ eV to 7.9% at E = $10^{20}$ eV, and to the uncertainty on the depth of the shower maximum ($X_{\max}$) from 3.3 g cm$^{-2}$ at E = $10^{17.5}$ eV to 7.3 g cm$^{-2}$ at E = $10^{20}$ eV. The phase function and wavelength dependence contribute 1% and 0.5% in energy and 2% and 0.5% in $X_{\max}$, respectively [2]. Clouds can distort the light profiles of showers, and give a significant contribution to the hybrid exposure of the detector and therefore to the hybrid spectrum.

The Central Laser Facility (CLF) [1] produces calibrated UV laser beams every 15 minutes during FD data acquisition from a position nearly equidistant from three out of four FDs. A similar facility, the eXtreme Laser Facility (XLF), was completed in November 2008, at a symmetric position with respect to the CLF and the FDs. Both systems can produce vertical and inclined laser beams at 355 nm, having a nominal energy of 7 mJ per pulse, which is approximately equivalent to the amount of fluorescence light produced by a $10^{20}$ eV shower. The number of photons reaching the FDs depends on the number of photons at the source and on the atmospheric conditions between the laser and the detector. Using the independently measured laser pulse energy, the aerosol transmission can be inferred. Clouds along the laser beam and between the laser site and the FDs can be identified.

Four elastic backscatter LIDAR stations are operating (the last one since May 2008). Each station has a fully steerable frame equipped with a UV laser, mirrors and PMTs for the detection of the elastic backscattered light. During FD data taking, hourly sets of scans are performed out of the FDs field of view to avoid interference with the FD telescopes to record local aerosol conditions and clouds. Elastic LIDARs also provide a rapid monitoring mode after the detection of events of particular physical interest, scanning very high-energy showers tracks within 10 minutes from detection (Shoot the Shower, StS) [5]. In addition, the Los Leones LIDAR station is equipped with a vertical Raman LIDAR system, detecting the inelastic Raman backscattered light. The molecular Raman cross section is small, therefore during Raman runs the laser is fired at high power to collect enough light. To avoid interference with the FD, Raman LIDAR runs are limited to 20 minutes at the beginning and 20 minutes at the end of the FD acquisition.

A Raytheon 2000B infrared cloud camera (IRCC) with a spectral range from 7 to 14 $\mu$m is located on the roof of each FD building to determine the cloud coverage. Each IRCC is housed within a weather protective box and is mounted on a pan-and-tilt device. During FD data acquisition, each IRCC takes a picture of the field of view of the 24 telescopes every 5 minutes, to flag pixels "covered" by clouds. In addition, a full sky scan is performed every 15 minutes to take a photograph of the entire sky above each FD site. A bi-dimensional map of the sky is produced.

Two techniques based on the analysis of FD background data recorded during acquisition have also been developed to retrieve cloud coverage information. Clouds can be identified by studying the changes in the brightness of the night UV sky, appearing as very dark patches against the bright night sky. The FD background data and IRCC analyses are complementary with the LIDAR and CLF studies that provide the height of cloud layers to achieve a better accuracy in cloud studies and obtain a 3-dimensional map of the sky.

## III. Aerosol Optical Depth Measurements

The light scattered out of the CLF beam produces tracks recorded by the FD telescopes. Laser light is attenuated in its travel towards FDs as the fluorescence light emitted by a shower. Therefore, the analysis of the amount of CLF light that reaches the FD building can be used to infer the attenuation due to aerosols, once the nominal energy is known. An hourly aerosol characterisation is provided in the FD field of view with two independent approaches using the same vertical laser events. The first method (Data Normalised Analysis) consists of an iterative procedure that compares hourly average profiles to reference profiles chosen in extremely clear (aerosol free) nights. The procedure starts with the definition of an average hourly profile obtained merging the corresponding four quarter-hour profiles.

A first estimate of $\tau_a(h)$ is given by:

$$\tau_a^{\text{first}}(h) = -\frac{\ln\left(I_{\text{hour}}(h)/I_{\text{aerfree}}(h)\right)}{1 + 1/\sin\theta}$$

where $I_{\text{hour}}$ is the average hourly laser profile, $I_{\text{aerfree}}$ is the reference profile and $\theta$ is the elevation angle of the laser track point at height $h$. This calculation does not take into account the laser beam scattering due to aerosols; to overcome this, $\tau_a^{\text{first}}(h)$ is differentiated to calculate the extinction $\alpha(\text{h})$ over short intervals in which the aerosol scattering conditions change slowly. Finally, $\tau_a(h)$ is estimated re-integrating $\alpha(h)$.

The second procedure (Laser Simulation Analysis) compares quarter-hour CLF profiles to simulated laser events generated varying over more than 1100 aerosol conditions to find the best compatibility. Aerosol-free profiles are used to fix the energy scale between simulations and real events. A parametric model of the aerosol attenuation is adopted, described by the Horizontal Attenuation Length ($\text{L}_{\text{mie}}$) and the Scale Height ($\text{H}_{\text{mie}}$) :

$$\tau_a(h) = -\frac{H_{mie}}{L_{mie}}\left[e^{\left(-\frac{h}{H_{mie}}\right)} - e^{\left(-\frac{h_0}{H_{mie}}\right)}\right]$$

where $\text{h}_0$ is the altitude above sea level of the detector.

Aerosol-free nights are needed as a reference in both analyses. A procedure to identify these extremely clear conditions in real data has been developed: the shape of each real profile is compared to the one of an aerosol-free simulated profile using a Kolmogorov test that checks their compatibility. The profile with the highest probability is chosen as the reference. Aerosol-free conditions occur more frequently during austral winter.

An example of the good agreement between a typical hourly vertical aerosol optical depth profiles measured with the Data Normalised and the Laser Simulation analyses is shown in figure 2.

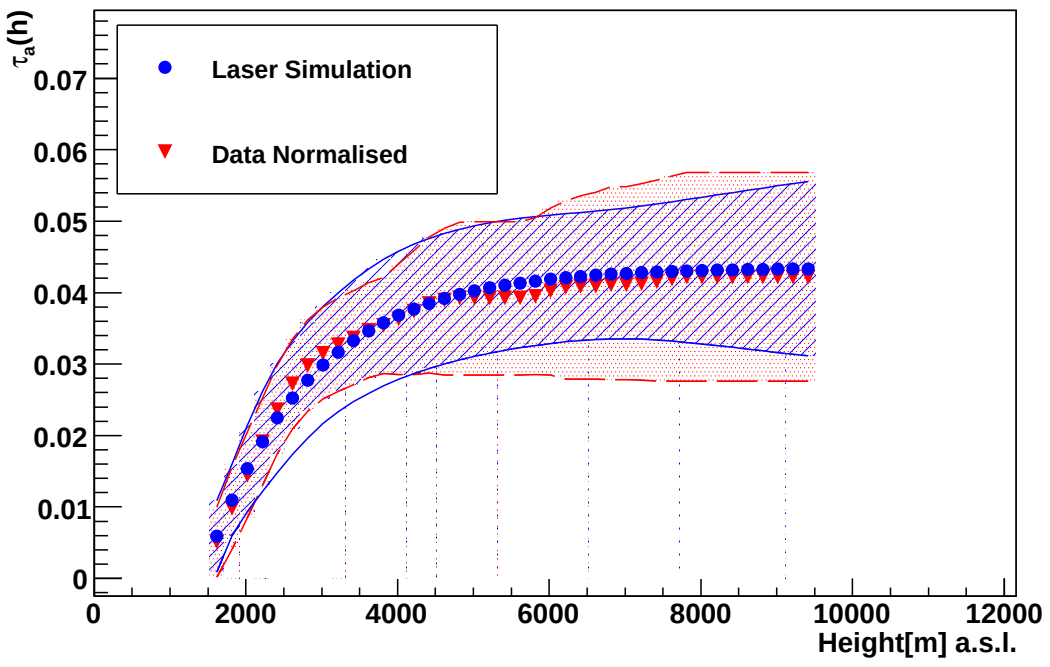

Fig. 2. Comparison of a $\tau_a(h)$ profile estimated by the Laser Simulation and the Data Normalised analyses

The results produced with these two independent analyses are fully compatible, as shown in fig. 3: the average $\tau_a$(3 km) above the detector in the period from January 2005 to December 2008 is 0.04 $\pm$ 0.01.

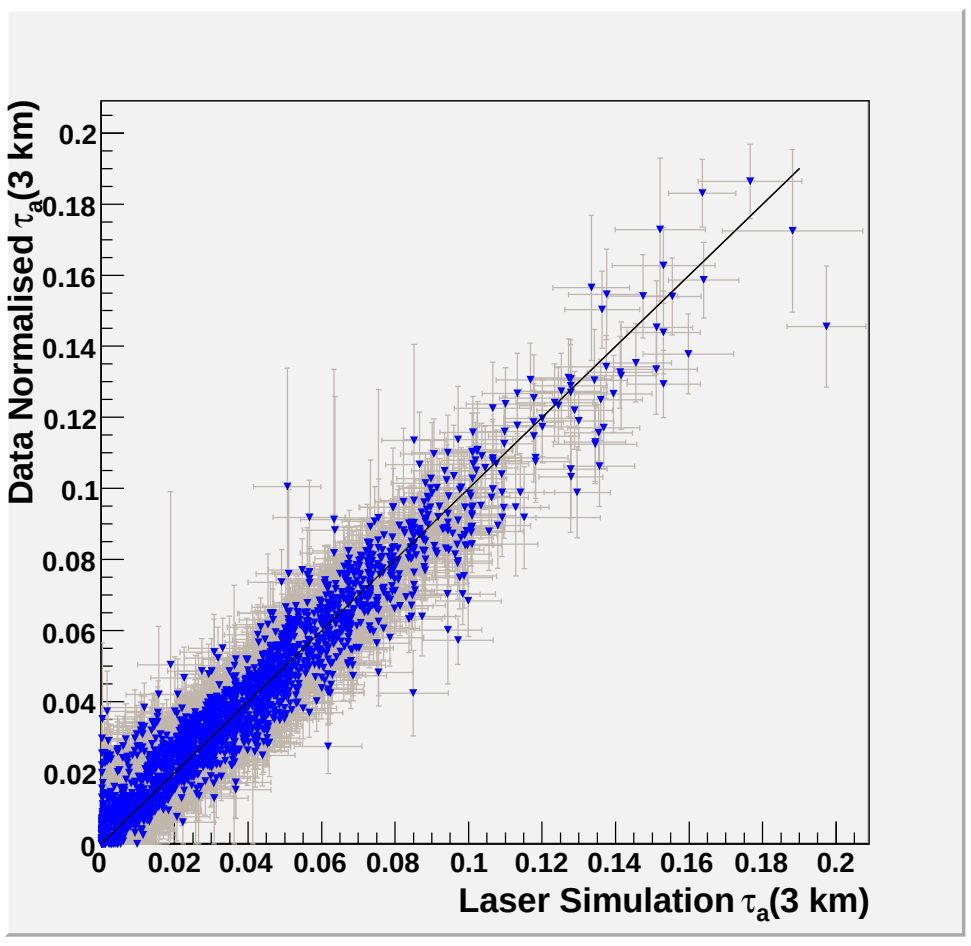

Fig. 3. $\tau_a(\mathrm{h})$ estimated by CLF analyses

By studying the vertical aerosol optical depth as a function of time, over a period of 4 years of data, a clear seasonal variation is observed, as shown in figure 4. Austral winter is the season with lower aerosol attenuation.

In addition to the CLF estimate of aerosol conditions, the four LIDAR stations provide a local estimate of $\tau_a(\mathrm{h})$ and $\alpha(\mathrm{h})$ using a multiangular inversion procedure [4]. Every hour, the LIDAR telescopes sweep the sky in a set pattern, pulsing the laser at 333 Hz and observing the backscattered light with the optical receivers. However, except for the StS mode [5] and a short hourly set of horizontal shots towards CLF, the LIDAR laser beams point outside the FD telescopes field of view to avoid triggering the detector.

## IV. Clouds Detection

Clouds have a significant impact on shower reconstruction, blocking the transmission of light in its travel from the emission point to the fluorescence telescopes, or enhancing the amount of light scattered towards the FD, depending on the position of the cloud itself.

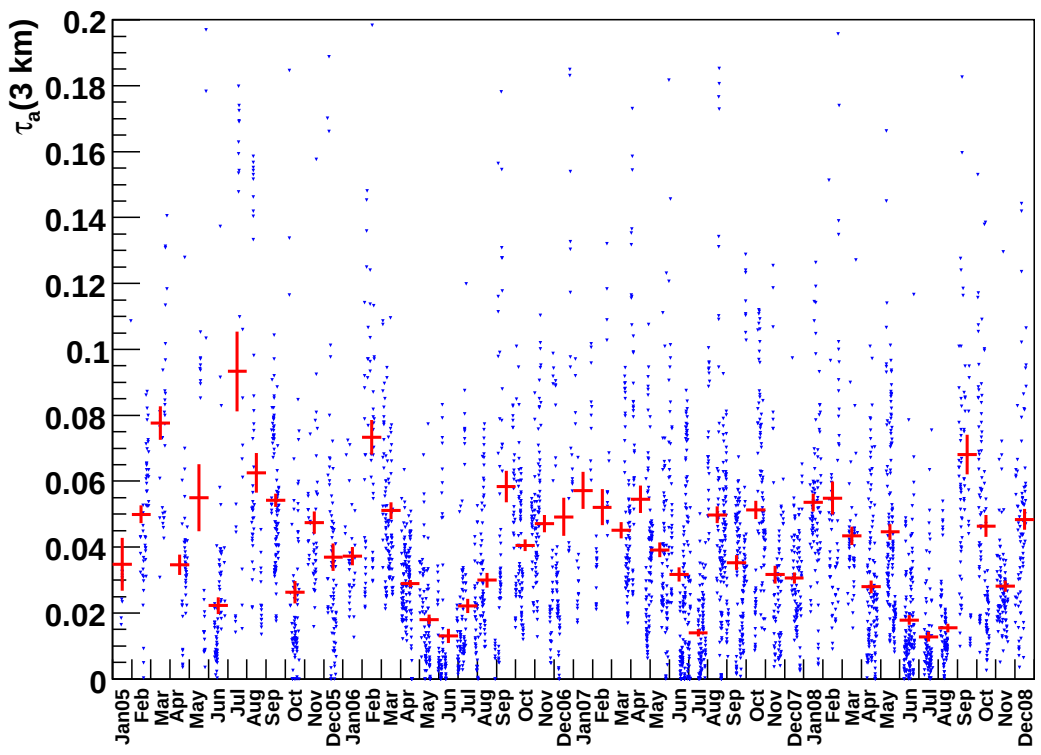

Fig. 4. $\tau_\mathrm{a}(3\mathrm{km})$ as a function of time. Lower values of $\tau_\mathrm{a}(3\mathrm{km})$ happen in austral winter (June - July)

The cloud coverage can be determined by analysing the FD background data: the variance of the baseline fluctuation is recorded every 30 s, providing a reasonable estimate of the changes in the brightness of the sky. As already mentioned, two approaches have been developed. “Star Visibility Method” : as stars are visible in the background data, it is possible to predict at what time a particular star would be visible. A null detection of the star indicates the presence of a cloud in the field of the viewing pixel. “Background Variation Method” : clouds are good absorbers of UV radiation, therefore they appear as dark areas against the bright background of the UV night sky. Sudden drops of the brightness of a part of the sky are an indication of an obscuring cloud. In fig. 5, an example of change in brightness from a single pixel during one night is shown: the peaks are stars crossing the field of view, while the drops are likely to be clouds.

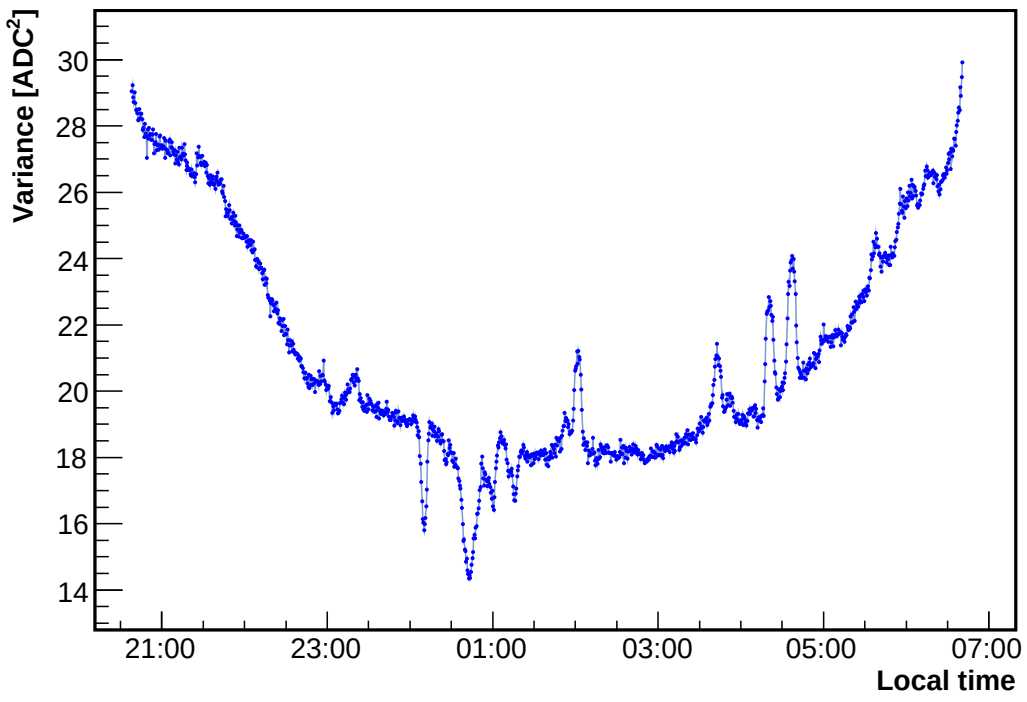

Fig. 5. Typical night sky background variation from one pixel

The four IRCCs record the cloud coverage making a photograph of the field of view of each telescope every 5 minutes during FD acquisition. The image data are processed and a coverage mask is created for each pixel of the telescope to identify cloud covered pixels to be removed from the shower reconstruction procedure. Cloud cameras are not radiometric, therefore each pixel value is proportional to the difference between the temperature in the viewing direction and the average temperature of the entire scene. In fig. 6, the raw IRCC images of the FD field of view are shown together with the final mask. The database is filled with the coverage for each pixel in the map.

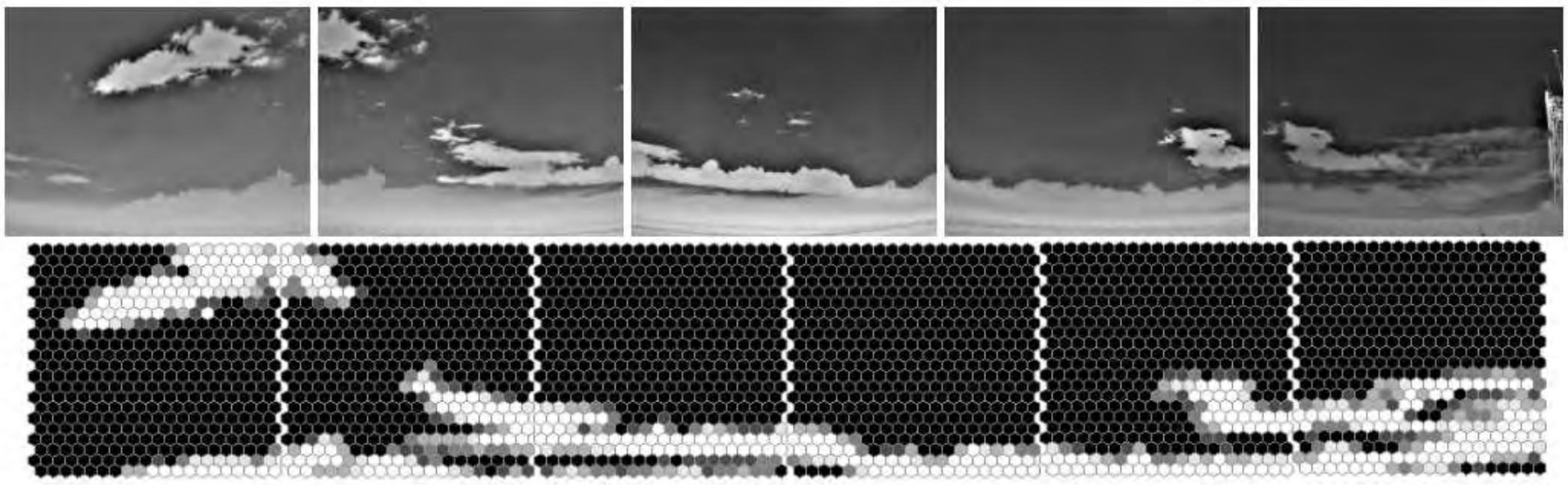

Fig. 6. Top: raw IRCC image. Bottom: FD pixels coverage mask: lighter values on the greyscale represent greater cloud coverage

While the IR cloud cameras and the FD background data analyses record the cloud coverage in the FD field of view, they cannot determine cloud heights, that must be measured using the LIDAR stations and CLF. In cloud detection mode, LIDAR telescopes sweep the sky with a continuous scan in two orthogonal paths with fixed azimuthal angle, one of which is along the central FD azimuth angle, with a maximum zenith angle of $45°$. The full scan takes 10 minutes per path. Clouds are detected as strong localised scattering sources, and the timing of the scattered light is related to the cloud height. The cloud finding algorithm starts with the subtraction of the expected signal for a simulated purely molecular atmosphere ($S_{mol}$) to the real one ($S_{real}$). The obtained signal is approximately constant before the cloud, and has a non-zero slope inside the cloud. A second-derivative method to identify cloud candidates and obtain cloud thickness is applied. LIDARs provide hourly information on cloud coverage and height.

In fig. 7, the intensity of the backscattered light as a function of height and horizontal distance from the LIDAR station is shown.

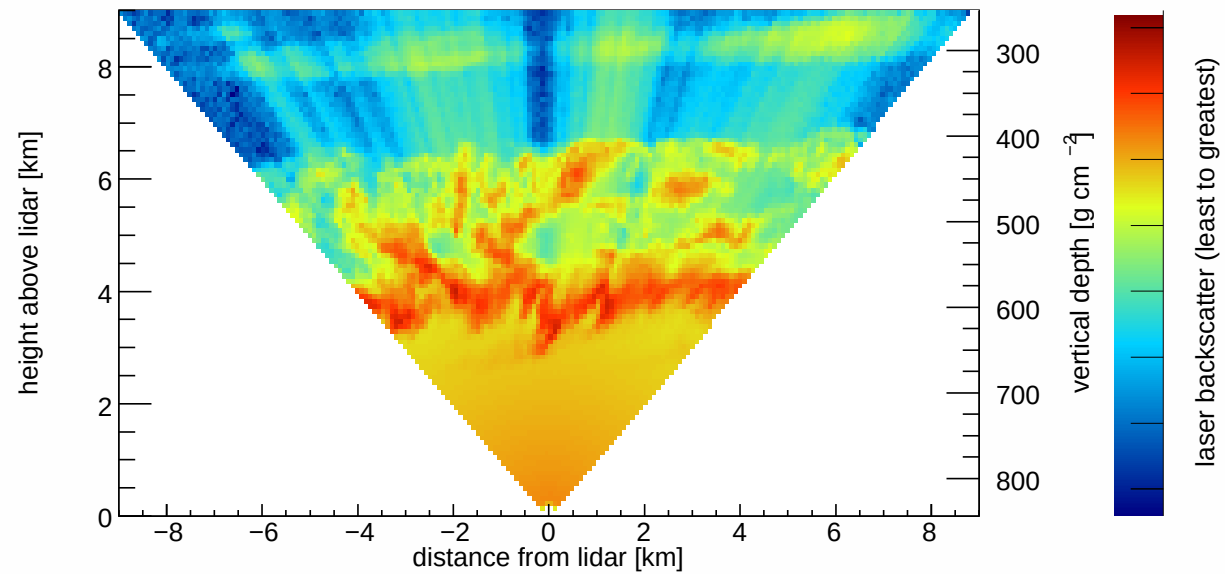

Fig. 7. A cloud layer around 3.5 km height as detected by the LIDAR

The Central Laser Facility and the eXtreme Laser Facility can be used to detect clouds along the vertical laser path and between their position and the FDs, looking at the profiles of photons collected at the FD buildings, since clouds can enhance or block the trasmitted light, depending of their position. A cloud positioned directly along the vertical laser track will scatter a greater amount of light in every direction, producing a peak in the light profile. In this case the cloud is directly above the laser facility site, and timing of the scattered light is related to the cloud height allowing to define the height of the cloud layer. If clouds are between the laser source and the FD, a local decrease in the laser light profile is observed. In this case the timing of the received light is not directly related to the cloud height, and only the cloud coverage in the FD field of view can be defined. A database is filled with the informations on the height of the observed cloud layers.

## V. Conclusions

The Pierre Auger Observatory operates a huge system to provide continuous measurements of the highly variable aerosol attenuation and for the detection of clouds, main sources of uncertainties in the shower reconstruction. The highest energy air showers are viewed at low elevation angles by the Fluorescence Detectors and through long distances in the lower part of the atmosphere, where aerosols are in higher concentration and therefore the aerosol attenuation becomes increasingly important. Also clouds have a significant effect on shower reconstruction. All the described instruments are operating, and most of the results are currently used in the reconstruction of shower events.

# Acknowledgements

The successful installation and commissioning of the Pierre Auger Observatory would not have been possible without the strong commitment and effort from the technical and administrative staff in Malargüe.

We are very grateful to the following agencies and organizations for financial support:

Comisión Nacional de Energía Atómica, Fundación Antorchas, Gobierno De La Provincia de Mendoza, Municipalidad de Malargüe, NDM Holdings and Valle Las Leñas, in gratitude for their continuing cooperation over land access, Argentina; the Australian Research Council; Conselho Nacional de Desenvolvimento Científico e Tecnológico (CNPq), Financiadora de Estudos e Projetos (FINEP), Fundação de Amparo à Pesquisa do Estado de Rio de Janeiro (FAPERJ), Fundação de Amparo à Pesquisa do Estado de São Paulo (FAPESP), Ministério de Ciência e Tecnologia (MCT), Brazil; AVCR AV0Z10100502 and AV0Z10100522, GAAV KJB300100801 and KJB100100904, MSMT-CR LA08016, LC527, 1M06002, and MSM0021620859, Czech Republic; Centre de Calcul IN2P3/CNRS, Centre National de la Recherche Scientifique (CNRS), Conseil Régional Ile-de-France, Département Physique Nucléaire et Corpusculaire (PNC-IN2P3/CNRS), Département Sciences de l'Univers (SDU-INSU/CNRS), France; Bundesministerium für Bildung und Forschung (BMBF), Deutsche Forschungsgemeinschaft (DFG), Finanzministerium Baden-Württemberg, Helmholtz-Gemeinschaft Deutscher Forschungszentren (HGF), Ministerium für Wissenschaft und Forschung, Nordrhein-Westfalen, Ministerium für Wissenschaft, Forschung und Kunst, Baden-Württemberg, Germany; Istituto Nazionale di Fisica Nucleare (INFN), Ministero dell'Istruzione, dell'Università e della Ricerca (MIUR), Italy; Consejo Nacional de Ciencia y Tecnología (CONACYT), Mexico; Ministerie van Onderwijs, Cultuur en Wetenschap, Nederlandse Organisatie voor Wetenschappelijk Onderzoek (NWO), Stichting voor Fundamenteel Onderzoek der Materie (FOM), Netherlands; Ministry of Science and Higher Education, Grant Nos. 1 P03 D 014 30, N202 090 31/0623, and PAP/218/2006, Poland; Fundação para a Ciência e a Tecnologia, Portugal; Ministry for Higher Education, Science, and Technology, Slovenian Research Agency, Slovenia; Comunidad de Madrid, Consejería de Educación de la Comunidad de Castilla La Mancha, FEDER funds, Ministerio de Ciencia e Innovación, Xunta de Galicia, Spain; Science and Technology Facilities Council, United Kingdom; Department of Energy, Contract No. DE-AC02-07CH11359, National Science Foundation, Grant No. 0450696, The Grainger Foundation USA; ALFA-EC / HELEN, European Union 6th Framework Program, Grant No. MEIF-CT-2005-025057, European Union 7th Framework Program, Grant No. PIEF-GA-2008-220240, and UNESCO.